\documentclass[epj]{svjour}
\usepackage{graphics}
\usepackage{epsfig}
\def\etal{{\it et al.}}
\begin{document}
\title{Experimental Results in Heavy Flavor Physics }
\author{Sheldon Stone
}                     
%
%
\institute{Department of Physics, Syracuse University, Syracuse, NY, USA, 13244-1130 \email{Stone@physics.syr.edu}}
\date{Received: October 15, 2003}
%
\abstract{The interplay of experiment and theory is explored in
the context of current data on $b$ and $c$ decay. Measurements
of $|V_{cb}|$ and $|V_{ub}|$ are extracted from existing data. Conservative estimates give
$\left|V_{cb}\right|=(42.4\pm 1.2_{exp} \pm 2.3_{thy})\times
10^{-3}$ and $|V_{ub}|=(3.90\pm 0.16_{exp}\pm 0.53_{thy}) \times 10^{-3}.$ Using these values along with data on $B_d$, $B_s$ mixing and CP violation in the $K_L$ system, the
allowed region of the CKM parameters $\rho$ and $\eta$ is derived.
Tests of factorization in two-body hadronic $B$ decays to one heavy and one light meson are shown and compared with modern theories which are also used to see if there is new physics in two-body $B$ decays to light mesons.
The two new narrow $D_{sJ}$ states, discovered by BaBar and CLEO,
respectively, are interpreted in light of the observation of these
states in $B$ decays by Belle.
} 
\maketitle
\section{Introduction}
\label{intro} Our physics goals include discovering, or helping to
interpret, New Physics found elsewhere using $b$ and $c$ decays.
We already know that there must be New Physics because the
Standard Model cannot explain the large observed Baryon Asymmetry
in the Universe or Dark Matter \cite{SM}. We also need to measure
Standard Model parameters, the ``fundamental constants" revealed
to us by studying Weak interactions. Furthermore, understanding
the theory of strong interactions, QCD, is necessary to interpret
our measurements.

A complete picture requires many studies including rare decays and
CP Violation; the latter is covered by H. Yamamoto
\cite{Hitoshi}. I will give an overview and cover rare decays, and
show how they can uncover new physics. Interpreting fundamental
quark decays requires theories or models than relate quarks to
hadrons in which they live and die. I will discuss some relevant
concerns. Theoretical issues are dealt with in more depth by T.
Mannel \cite{Mannel}. Yamamoto also covers future experiments.

\section{Lifetimes}
\label{sec:lifetimes}

\begin{figure}
\centerline{\epsfig{figure=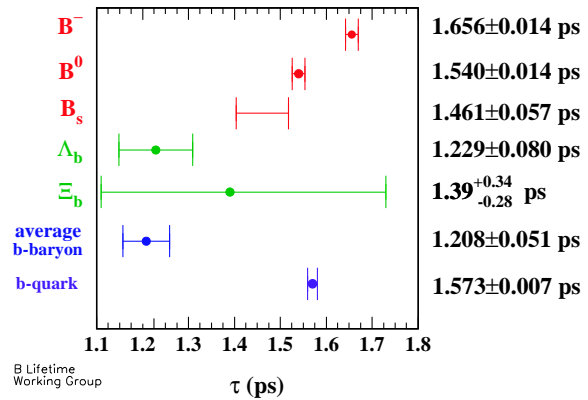,height=2.1in}}
\caption{Current measurement of lifetimes of $b$-flavored
hadrons.}
\label{B_life}       
\end{figure}

Lifetimes ($\tau$) set the width ($\Gamma$) scale since
$\tau\cdot\Gamma=\hbar$. Lifetimes of $b$-flavored hadrons are
shown in Fig.~\ref{B_life} as compiled by the B-lifetime working
group \cite{blife}. Note that the ratio of $B^+$ to $B^{\rm o}$
lifetimes is 1.073$\pm$0.014, a 5.2$\sigma$ difference, while the
$\Lambda_b$ and $b$-baryon lifetimes are lower than the $B^{\rm
o}$.

\begin{figure}
\centerline{\epsfig{figure= 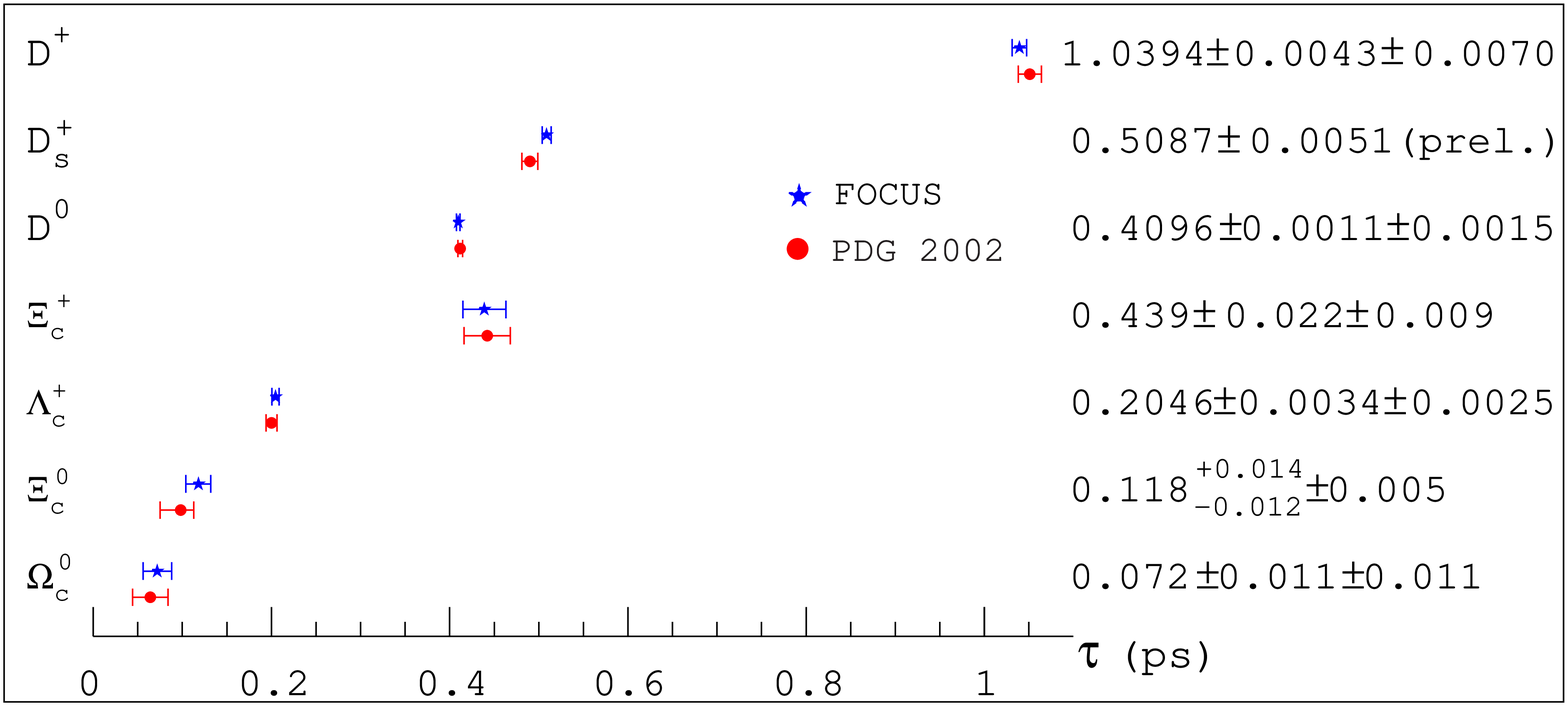,height=1.5in}}
\caption{New FOCUS measurement of charm lifetimes listed and compared
on the plot with the PDG 2002 values.}
\label{daniele}       
\end{figure}

New charm lifetimes from FOCUS \cite{pedrini} are very precisely
measured. See Fig.~\ref{daniele}.

\section{The Basics: Quark Mixing and the CKM Matrix}
\label{CKM} The CKM matrix parameterizes the mixing between the
mass eigenstates and weak eigenstates as couplings between the
charge +2/3 and -1/3 quarks. I use here the Wolfenstein
approximation \cite{Wolf} good to order $\lambda^3$ in the real
part and $\lambda^4$ in the imaginary part: $V_{CKM}=$ \small{
\begin{equation}
\label{eq:CKM}
\left(\begin{array}{ccc}
1-\lambda^2/2 &  \lambda & A\lambda^3(\rho-i\eta)(1-\lambda^2/2) \\
-\lambda &  1-\lambda^2/2-i\eta A^2\lambda^4 & A\lambda^2(1+i\eta\lambda^2) \\
A\lambda^3(1-\rho-i\eta) &  -A\lambda^2& 1
\end{array}\right).
\end{equation}}\normalsize

In the Standard Model $A$, $\lambda$, $\rho$ and $\eta$ are
fundamental constants of nature like $G$, or $\alpha_{EM}$; $\eta$
multiplies $i$ and is responsible for all Standard Model CP
violation. We know $\lambda$=0.22, $A \sim$0.8 and we have
constraints on $\rho$ and $\eta$.

\begin{figure}
\centerline{\epsfig{figure= 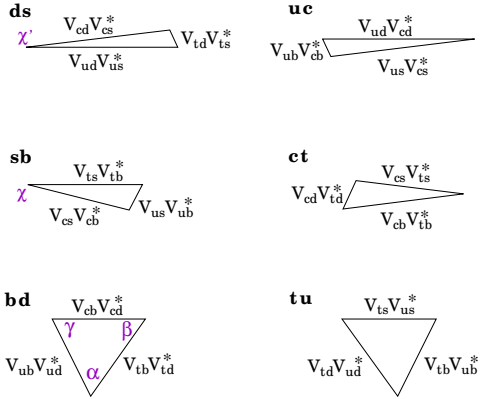,height=2.1in}}
\caption{The 6 CKM triangles resulting from applying unitarity constraints
to the indicated row and column. The CP violating angles are also shown.}
\label{six_tri}       
\end{figure}

Applying unitarity constraints allows us to construct the six
independent triangles shown in Fig.~\ref{six_tri}. Another basis
for the CKM matrix are four angles labeled as $\chi$, $\chi'$ and
any two of $\alpha$, $\beta$ and $\gamma$ since $\alpha +\beta
+\gamma =\pi$ \cite{akl}.

\begin{figure}
\centerline{\epsfig{figure=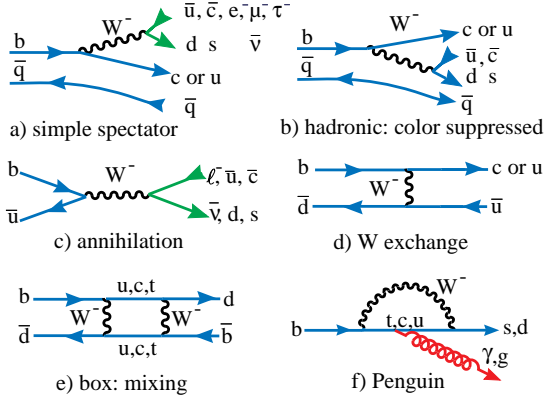,height=2.1in}}
\caption{Some $B$ decay diagrams.}
\label{Bdiagrams2}       
\end{figure}

$B$ meson decays can occur through various processes. Some decay
diagrams are shown in Fig.~\ref{Bdiagrams2}. The simple spectator
diagram is dominant. Semileptonic decays, which proceed through
this diagram, are very useful and are discussed next.

\section{The Semileptonic Branching Ratio}
\label{sec:bsemi} 
\begin{figure}
\centerline{\epsfig{figure=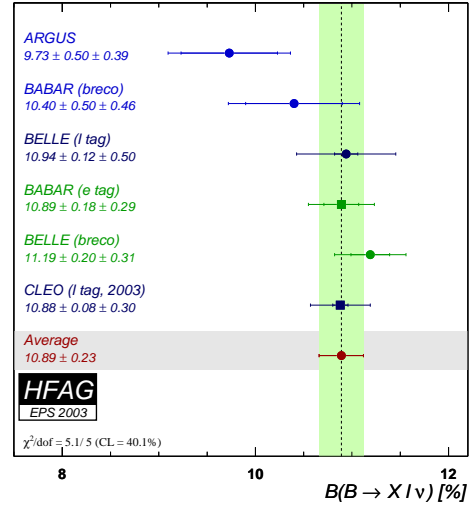,height=2.7in}}
\caption{$B$ semileptonic branching ratio measurements from $\Upsilon(4S)$ decays.}
\label{Y4s_bsl}       
\end{figure}

The total semileptonic branching of $B$ mesons
(${\cal{B}}_{sl}$) can be measured using the process
$\overline{B}\to X \ell^-\overline{\nu}$ in $\Upsilon(4S)$ decay,
that contains an almost equal mixture of pair-produced charged and
neutral $B$ mesons. One problem arises because the decay sequence
$B\to D\to Y\ell^+\nu$ also produces leptons, albeit of lower
momentum. The charge of these leptons, however, is opposite to
those of the ones produced directly by the $B$ decay. ARGUS long
ago developed a technique of tagging the flavor of one $B$ using a
high momentum lepton. This allows the specification of the charge
of the lepton directly from the second $B$ in the pair.
Corrections must be made for $B^{\rm o}-\overline{B}^{\rm o}$
mixing and also leptons produced from $D_s$ or $D$ decays when
they are produced by the virtual $W^-$ as in
Fig.~\ref{Bdiagrams2}(a). Fig.~\ref{Y4s_bsl} shows the relevant
measurements that result in ${\cal{B}}_{sl}$=10.89$\pm$0.23\%
\cite{HFAG}.

LEP measurements average 10.59$\pm$0.22\%. This number after
correction for other $b$-species, by using the measured lifetimes,
under the assumption that the semileptonic widths of all $b$
species are equal, becomes 10.76$\pm$0.22\% for an average of
$B^{\rm o}$ and $B^-$, in excellent agreement with the
$\Upsilon(4S)$ measurements.

\section{Determination of \boldmath $|V_{cb}|$ and $|V_{ub}|$}
\label{sec:VcbVub}
\subsection{Introduction: Theory versus Models}
\label{subsec:theory}
Theories describe phenomena and make predictions based on general principles. They can have one or
two unknown parameters (e.g. coupling constants) and if not exact, must prescribe a convergent
series approximation. Some examples are Lattice QCD (unquenched) and Heavy Quark Effective Theory
(HQET).

Models contain assumptions. It is not only that the models may be wrong that causes us a problem,
just as serious is that the errors on the predictions are difficult to estimate.

\subsection{\boldmath $|V_{cb}|$}
\label{subsec:Vcb}
\subsubsection{\boldmath Using Exclusive $B\to D^*\ell^-\bar{\nu}$ Decays}
We are in the fortunate situation here of having a theory, Heavy
Quark Effective Theory (HQET) formulated by M. Wise and the late
N. Isgur \cite{Isgur-Wise}. This theory is based on the idea that
QCD is flavor independent, so in the limit of infinitely heavy
quarks the transition $q_a\to q_b$ occurs with unit form factor,
($F(1)=1$) when the quarks are moving with the same invariant
four-velocity, $\omega$. Corrections to $F(1)$ for the fact that
the $b$ and $c$ are not infinitely heavy are calculable in terms
of a series, $\sum_{n}C_n(1/m_{i,j})^n$, where $i$ and $j$ refer to $b$
and $c$, along with QCD corrections.

For determining $|V_{cb}|$ it is best to use the reaction $\overline{B}\to
D^*\ell^-\bar{\nu}$, because it has a large branching rate and the
$1/m_{i,j}$ corrections vanish \cite{Luke}. Although, in principle
there are three independent form-factors for this decay, due to
the three possible $D^*$ spin states, in HQET they are all related
to one universal shape that can be measured. The idea is to
determine the decay rate at $\omega$ of 1. Here the $D^*$ is at
rest in the $B$ frame.

\begin{figure}
\centerline{\epsfig{figure=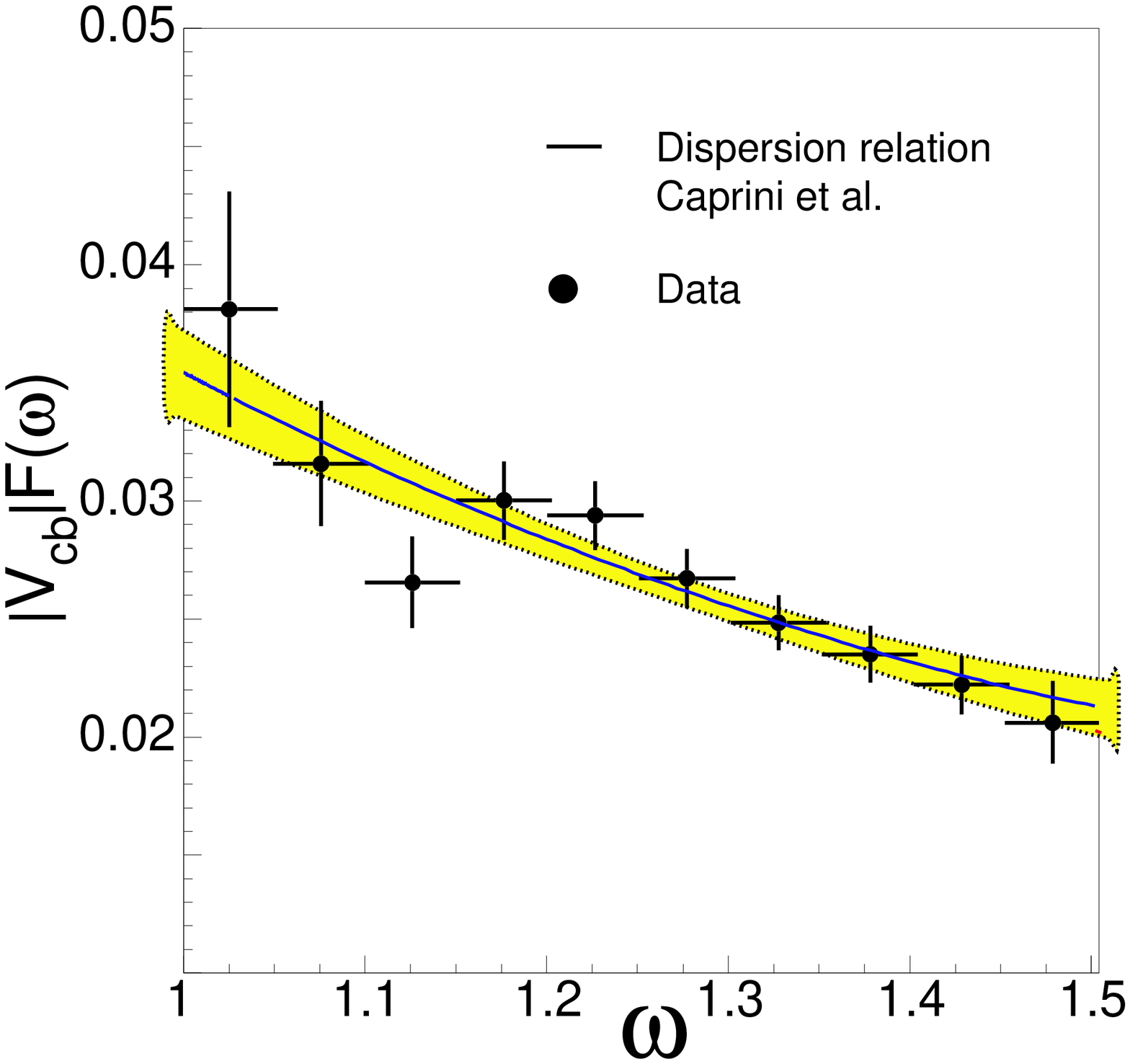,height=2.4in}}
\caption{Belle measurement of $F(\omega)|V_{cb}$ using
$\overline{B}^{\rm o}\to D^{*+}\ell^-\bar{\nu}$.}
\label{VcbF1}       
\end{figure}

\begin{figure}
\vspace{-.1cm}
\centerline{\epsfig{figure=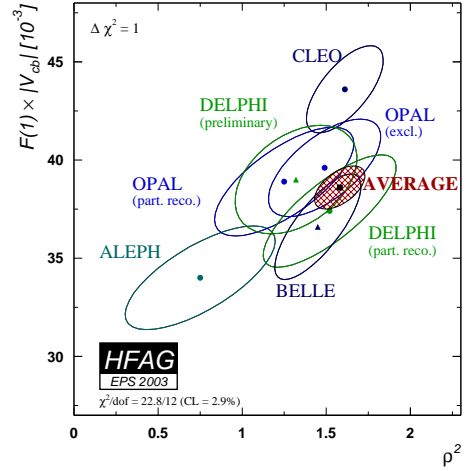,height=2.7in}}
\vspace{-.3cm} \caption{Compilation of $F(1)|V_{cb}|$ versus
$\rho^2$ measurements.}
\label{vcbf1vsrho2}       
\end{figure}

This measurement has been performed by several groups.
Fig.~\ref{VcbF1} shows recent measurements from Belle for the
reaction $\overline{B}^{\rm o}\to D^{*+}\ell^-\bar{\nu}$. To find
the value for $F(1)|V_{cb}|$ the data are plotted as a function of
$\omega$ and then fit to a shape function given by Caprini \etal~
\cite{Caprini}. The curvature of this function is denoted as
$\rho^2$ and also is found in the fit. Data from other experiments
are summarized in Fig.~\ref{vcbf1vsrho2}. All use the same reaction
except CLEO which also uses ${B}^-\to D^{*o}\ell^-\bar{\nu}$.

The world average values given by the Heavy Flavor Averaging Group
(HFAG) are $F(1)|V_{cb}|= (38.6\pm 1.1)\times 10^{-3}$ and
$\rho^2=1.58\pm 0.15$ \cite{HFAG}. To find $|V_{cb}$ we must
evaluate $F(1)$, where
$F(1)=\eta_{QED}\eta_{QCD}\left(1+\delta_{1/m^2}+...\right)$.
Current evaluations have $\eta_{QED}=1.007$, $\eta_{QCD}=0.960\pm
0.007$ at the two loop level. The $\delta_{1/m^2}$ term evaluation
has been summarized for the PDG by Artuso and Barberio
\cite{ArtusoB} giving $F(1)=0.91\pm 0.05$; eventually unquenched
Lattice Gauge calculations should be used. Thus we have
\begin{equation}
\left|V_{cb}\right|=(42.4\pm 1.2_{exp} \pm 2.3_{thy})\times
10^{-3}~~.
\end{equation}
\subsubsection{Using Inclusive Semileptonic Decays}
The Heavy Quark Expansion (HQE) is a framework that allows
predictions of the total widths, $b\to c$ or $b\to u$ semileptonic
widths. It relies on $m_b >> \Lambda_{QCD}$ and uses the Operator
Product Expansion to express the decay widths in a double series
in $1/m_q^n$ and $\alpha_s^n$ \cite{HQE}.

The HQE suffers from some serious problems. Terms of the order of
$1/m_b^3$ are multiplied by unknown functions, that it make it
difficult to evaluate the error at this order. Even more
importantly, there is inherent assumption, called ``Duality,"
having the meaning that integrated over enough phase space the
physical finite exclusive charm bound states and the inclusive
hadronic result will match at the quark level. However, there is
no known way to evaluate the error due to this assumption.

Using the HQE, Bigi predicts that the $\Lambda_b$ and $b$-baryon
lifetimes will be no more that 10\% lower than the $B^{\rm o}$
lifetime \cite{Bigi-LB}. However, $\tau_{\Lambda_b}$  and
$\tau_{b-{\rm baryon}}$ are lower by 20$\pm$5\% and 18$\pm$3\%,
respectively. It is possible, that that semileptonic widths are
easier to predict than hadronic widths, so this failure does not
necessarily make the HQE model useless, yet it does not add to our
confidence in using it.

What is required is an experimental test in semileptonic decays
that can be used to evaluate the errors due to duality and the
$1/m_b^3$ terms. Two tests are possible. One is to use the model
in charm decays. This hasn't been done since $m_c$ has been
considered as too small, but it could be done to see how much the
model diverged by measuring $V_{cs}$ and $V_{cd}$. Another test is
$V_{cb}$. Then we would have a pretty good measure of the
uncertainties for $V_{ub}$.

The HQE expansion uses three experimental parameters to describe
the decay rate to order $\left(\Lambda_{QCD}/m_b\right)^2$. First is
the kinetic energy of the residual $b$-quark motion,
\begin{equation}
\lambda_1={M_B\over 2} \langle B(v)\left|h_v(iD)^2
h_v\right|B(v)\rangle~.
\end{equation}
The second parameter is the chromo-magnetic coupling of the $b$-quark spin to the gluon
field, given by the measured $B^*-B$ splitting as 0.12 GeV$^2$,
\begin{equation}
\lambda_2=-{M_B\over 2} \langle
B(v)\left|h_v(g/2)\sigma^{\mu\nu}G_{\mu\nu}h_v\right|B(v)\rangle~.
\end{equation}
Finally the parameter $\overline{\Lambda}$ relates the meson and
quark masses via
\begin{eqnarray}
M_B&=&m_b+\overline{\Lambda}-(\lambda_1+3\lambda_2)/(2m_b)
\nonumber\\
M_{B^*}&=&m_b+\overline{\Lambda}-(\lambda_1-\lambda_2)/(2m_b)~.
\end{eqnarray}

\begin{figure}
\centerline{\epsfig{figure=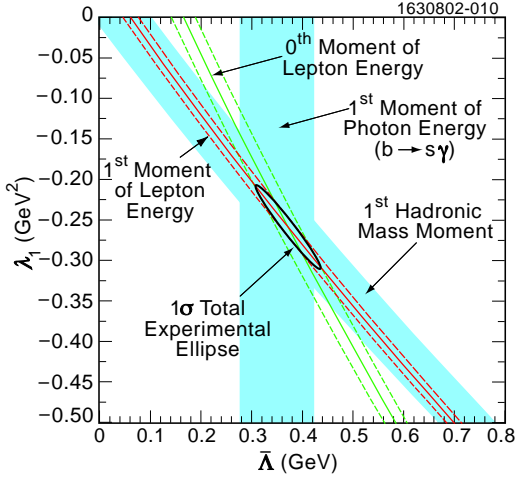,height=2.5in}}
\caption{Constraints on $\lambda_1$ versus $\overline{\Lambda}$
from CLEO hadronic and lepton energy moments and the first moment
of the photon energy in $b\to s\gamma$.}
\label{moments_all}       
\end{figure}

Both $\lambda_1$ and $\overline{\Lambda}$ can be determined by
measuring average quantities called ``moments." Useful terms in
$b\to c\ell\bar{\nu}$ decays are the average mass of the charmed
system decays, the zeroth and first moment of the lepton energy,
and in $b\to s\gamma$ decays the first moment of the photon
energy. The CLEO data are shown in Fig.~\ref{moments_all}
\cite{anal:HQE}.

From these measurements CLEO extracts $|V_{cb}|=(40.8\pm 0.5\pm
0.4 \pm 0.9)\times 10^{-3}$, where the first error is due to the
uncertainty on the total semileptonic width, the second error is
due to the uncertainty on the determination of $\lambda_1$ and
$\overline{\Lambda}$ and the third error reflects the theoretical
uncertainty. A similar value was found by Bauer \etal ~using these
and other data: $|V_{cb}| =(40.8 \pm 0.9)\times 10^{-3}$
\cite{BLLM}. A value within $\sim$7\% of the one found using $B\to
D^*\ell\nu$.

\begin{figure}
\centerline{\epsfig{figure=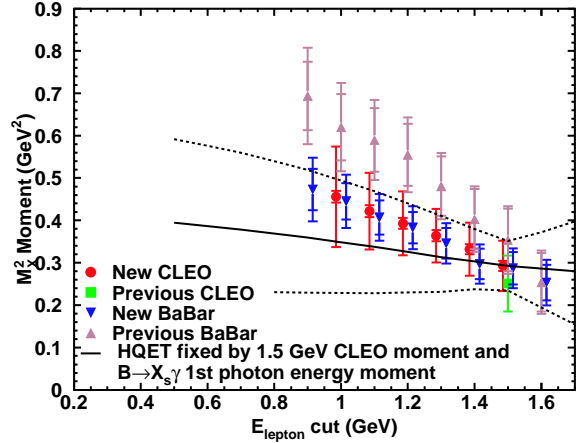,height=2.5in}}
\caption{First moment of the hadronic mass from BaBar and CLEO, both old and updated measurements.}
\label{Elliot}       
\end{figure}

Other groups have also done moment analyses. BaBar analyzes only
the hadronic moments. Their results are listed in
Table~\ref{tab:mom} \cite{babar:mom}. Fig.~\ref{Elliot} shows the
first hadronic moment as a function of minimum lepton momentum for
both old and updated BaBar and CLEO data. The data are in good
agreement and consistent with theory.

\begin{figure}
\centerline{\epsfig{figure=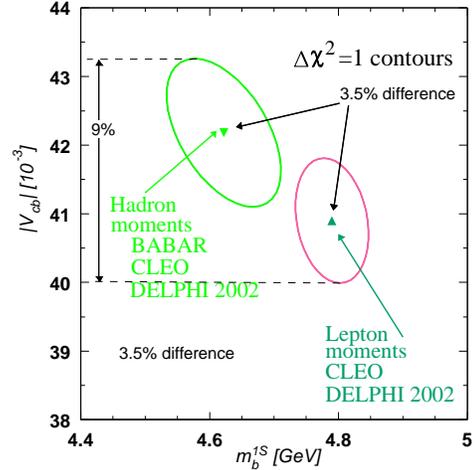,height=2.5in}}
\caption{Differences between hadron and lepton moments with experimental errors only.}
\label{vcb_vs_m_ss}       
\end{figure}

BaBar has also combined
other measurements of lepton and hadron moments separately and
find the differences shown in Fig.~\ref{vcb_vs_m_ss}. These
differences could indicate a duality violation, but the data need
to get better to establish that. Note that a difference of 0.2 GeV
in $m_b$ leads to a 20\% change in $V_{ub}$ using HQE. DELPHI have
also presented a new analysis where they use $m_b$ as an input
which gives them some sensitivity to the $1/m_b^3$ terms
\cite{DELPHI:mom}. To sum up, the situation is still evolving.

\begin{table}
\caption{Results of Moments Analyses}
\label{tab:mom}       
\begin{center}
\begin{tabular}{lccc}
\hline\noalign{\smallskip}
Group & $|V_{cb}|\times 10^{-3}$ & $m_b$ (GeV) \\
\noalign{\smallskip}\hline\noalign{\smallskip}
CLEO& $40.8\pm 0.6\pm 0.9$&$4.82\pm 0.07\pm 0.11$\\
BaBar &$42.1\pm1.0\pm0.7$  & $4.64\pm 0.09\pm 0.09$\\
DELPHI & $42.4\pm 0.6\pm 0.9$& input \\
\noalign{\smallskip}\hline
\end{tabular}
\end{center}
\end{table}

\subsection{\boldmath $|V_{ub}|$}
In this case there is no good theory. Modeling errors will
dominate the experimental errors and our path through this
discussion will be perilous.

\subsubsection{Using Exclusive Semileptonic Decays}
\begin{figure}
\centerline{\epsfig{figure=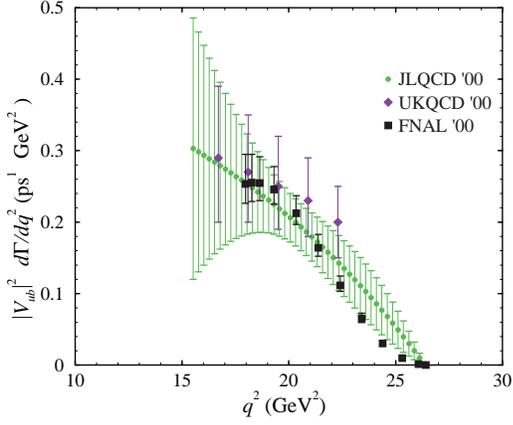,height=2.2in}}
\caption{Predictions of Lattice QCD for $B\to\pi\ell\nu$ from
\cite{Kronfeld}.}
\label{btopi_kronfeld}       
\end{figure}

The decays $B\to \pi (\rm ~or~\rho)\ell\nu$ can be used along with
predictions from Lattice QCD. Fig.~\ref{btopi_kronfeld} shows the
predictions of three quenched Lattice QCD calculations for
four-momentum transfer $q^2> 16$ GeV$^2$ from Kronfeld
\cite{Kronfeld}; the error currently on these calculations is
$\sim10$\% with an addition $\sim20$\% quenching error. There are
also calculations over the entire $q^2$ range from other models,
such as QCD sum rules \cite{Ball}.

\begin{figure}[htb]
\centerline{\epsfig{figure= 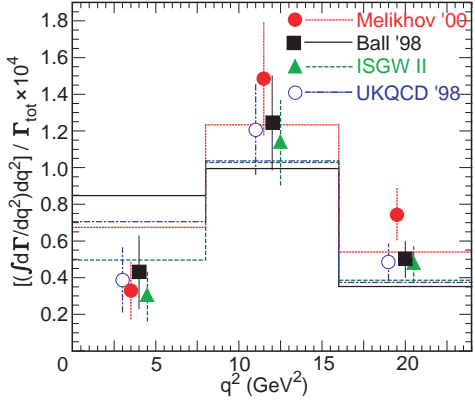,height=2.2in}}
\caption{CLEO measurements of the $q^2$ distribution for $B\to\rho\ell\nu$ compared with various models.}
\label{CLEO_Btorholnu}       
\end{figure}

\begin{figure} [hbt]
\centerline{\epsfig{figure=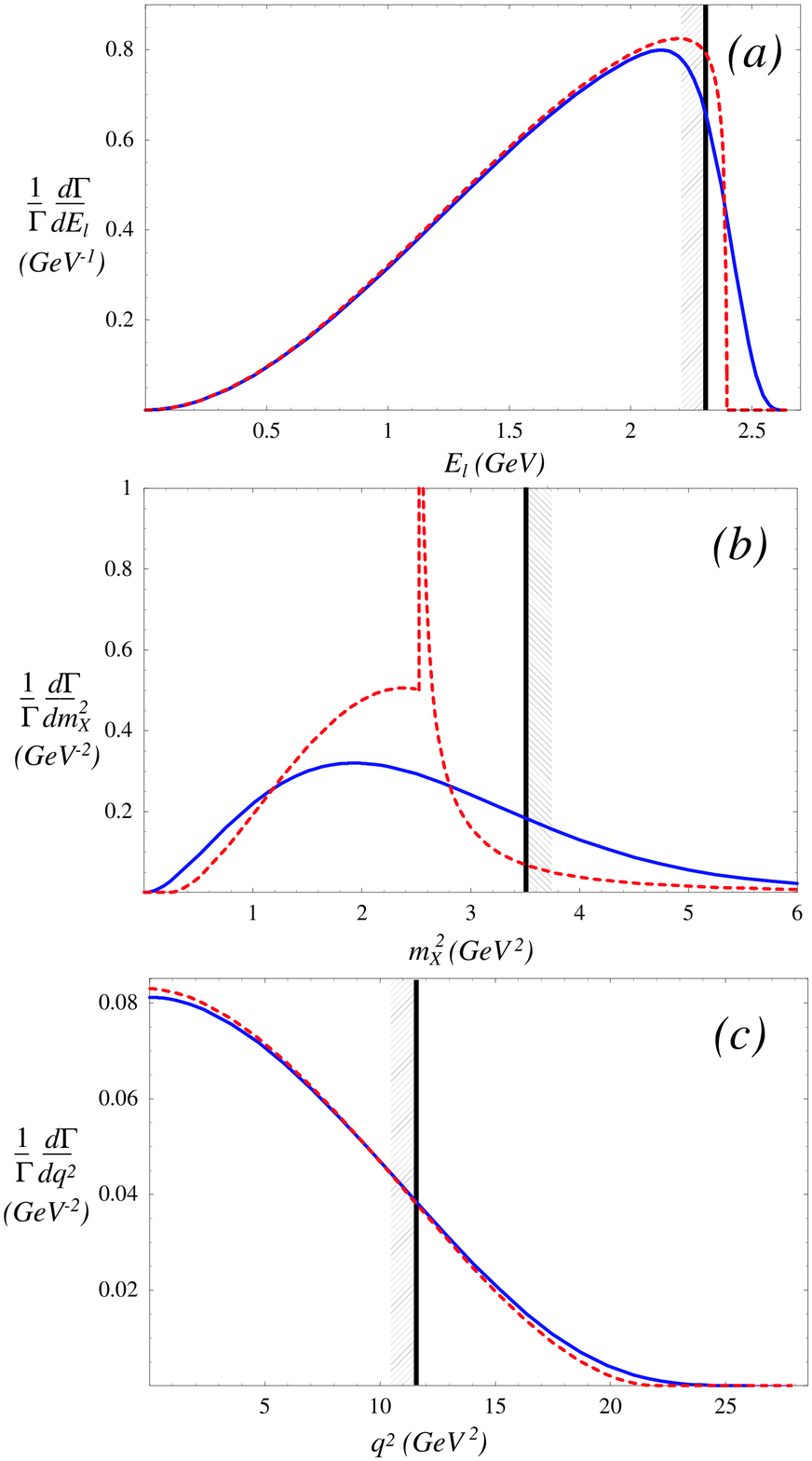,height=4.4in}}
\caption{The shapes of the lepton energy (a), hadronic mass (b)
and leptonic mass (c) spectra. The dashed curves are for
free $b$ quark decay while the solid curves have the Fermi motion
included in a model dependent manner. The unshaded side of the
vertical bars correspond to regions used by experiments to
suppress $b\to c$ background. (From Luke \cite{Luke_Vub}.)}
\label{threespectra}       
\end{figure}

CLEO made the first measurements of $B\to \pi (\rm
~or~\rho)\ell\nu$ decay rates and measured a rough $q^2$
distribution \cite{CLEO_vub_excl} that allows them to
significantly reduce the systematic errors in their efficiencies
(see Fig.~\ref{CLEO_Btorholnu}). Using quenched Lattice results
for $q^2>16$ GeV$^2$ and light cone sum rules for smaller $q^2$,
they quote
\begin{equation}
|V_{ub}|=(3.17\pm 0.17^{+0.16+0.53}_{-0.17-0.39}\pm 0.03)\times
10^{-3}~,
\end{equation}
where the errors are statistical, systematic, theoretical and
$\rho\ell\nu$ form-factor, respectively.

BaBar has also measured $B\to \rho\ell\nu$. Averaging
over one Lattice model and various form-factor models they quote
\begin{equation}
|V_{ub}|=(3.64\pm 0.22 \pm 0.25^{+0.39}_{-0.56})\times 10^{-3}~,
\end{equation}
with the same error sequence except without the last term \cite{BaBar_Vub_excl}.
The theoretical errors are assigned by each experiment. Are they large enough?

\subsubsection{Using Inclusive Semileptonic Decays}

Using the HQE framework a theoretical accuracy of $\sim$9\% is
obtainable if the entire $b\to u$ semileptonic decay rate was
measured. The theoretical limitations are imposed by the accuracy
on $m_b$ of about 0.1 GeV (the decay rate goes as $m_b^5$) and the
duality error which I limited to 7\% based on $|V_{cb}|$. However,
experimental cuts are required to reduce the $\sim$100 times larger
$b\to c$ rate and this usually means severely restricting the
phase space. Unfortunately, as emphasized by Luke \cite{Luke_Vub},
these cuts exaggerate the theoretical errors.
Fig.~\ref{threespectra} show the parton model rates and the
modification due to the Fermi motion of the $b$ quark for a
particular choice of spectral function $f(k^+)$. In general
$f(k^+)$ is unknown, although knowledge about it can be derived in
leading order only from the photon spectrum in $b\to s\gamma$.

Another pernicious consideration is that charged $B$ mesons can
have the $b$ and $\bar{u}$ quarks annihilate producing a
lepton-$\bar{\nu}$ pair along with two gluons, thereby breaking
the helicity suppression. The two gluons can turn into light
hadrons. This is, in fact, a $b\to u$ process but not the one we
want to consider because it supplements the $b\to u W^-$ rate, but
is not accounted for by the theoretical calculations. Estimating
this annihilation rate is difficult. A guess gives 3\%.
\cite{LLW-WA}.

Luke has summarized the additional sources of errors arising from
specific cuts restricting the signal phase space. The lepton
energy cut (Fig.~\ref{threespectra} (a)) is sensitive to $f(k^+)$
and sub-leading corrections, weak annihilation and may be more
sensitive to duality because only $\sim$10\% of the $b\to u$ phase
space is being used. The hadron mass cut (Fig.~\ref{threespectra}
(b)) is very sensitive to $f(k^+)$ and sub-leading corrections
(see \cite{Defazio}). Making both $q^2$ and hadronic mass cuts is
preferred if the minimum $q^2$ is as low as 7 GeV$^2$, although
there may be an increased sensitivity to $m_b$ here
\cite{Neub_mb}.

\begin{figure}
\centerline{\epsfig{figure=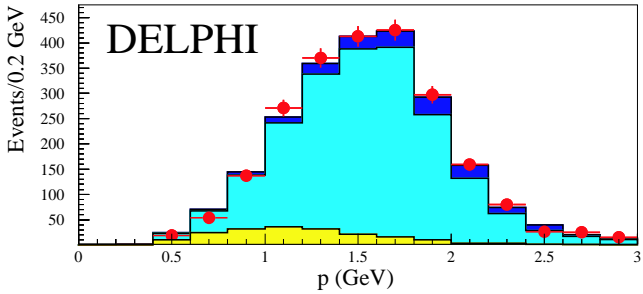,height=1.5in}}
\caption{The lepton momentum distribution in the $B$ rest frame
for $b\to u$ enriched decays. The data are the points, the dark
shaded region the $b\to u\ell\nu$ signal, the medium shaded region
the $b\to c\ell\nu$ background and the light shaded region other
backgrounds.}
\label{DELPHI_Vub}       
\end{figure}

Let us view the experimental results. ALEPH, DELPHI  and L3
select samples of charm-poor semileptonic decays using a large
number of selection criteria and performing a hadron mass cut $< 1.6$ GeV
\cite{LEP-Vub}. The DELPHI signal is shown as a function of lepton
momentum in Fig.~\ref{DELPHI_Vub}.

An average of all three measurements \cite{LEP-Vub-sum} gives the
value
\begin{equation}
|V_{ub}|=(4.04^{+0.41+0.43+0.24}_{-0.46-0.48-0.25}\pm 0.19)\times
10^{-3}~,
\end{equation}
where the first error reflects statistical and detector
systematics, the second from $b\to c$ modeling, the third $b\to u$
modeling and the fourth OPE uncertainties.

\begin{figure}[htb]
\centerline{\epsfig{figure=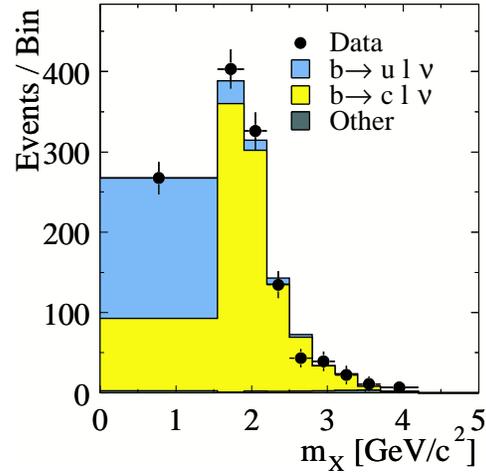,height=2.5in}}
\caption{The hadronic mass distribution from fully reconstructed tags from BaBar.}
\label{BaBar_Vub}       
\end{figure}

BaBar uses fully reconstructed hadronic $B$ decays and then look
for the semileptonic decay of the other $B$. They obtain an
excellent signal to background of 2.5 to 1. Their sample of $b\to
u\ell\nu$ events is plotted as a function of hadronic mass
Fig.~\ref{BaBar_Vub}. Using the sample in the lowest bin, they
quote
\begin{equation}
|V_{ub}|=(4.62\pm 0.28\pm 0.27 \pm 0.40 \pm 0.26)\times
10^{-3}~,
\end{equation}
where the errors are statistical, systematic, the fraction of $b\to u$ within their cuts and theory.

\begin{figure}[htb]
\centerline{\epsfig{figure=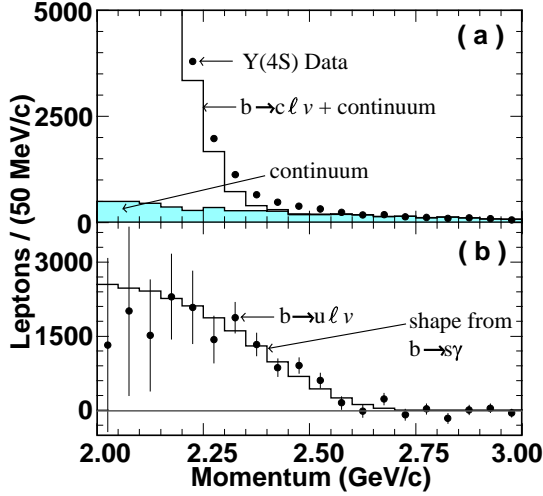,height=2.9in}}
\vspace{-0.2in}
\caption{The lepton momentum spectrum in the endpoint region from CLEO. (a) The data and (b) after continuum subtraction fit to a theoretical shape derived from the photon spectrum in $b\to s\gamma$.}
\label{CLEO_Vub}       
\end{figure}

Next are the measurements using the end of the lepton momentum spectrum beyond were $B \to D\ell\nu$ can be produced. CLEO pioneered this technique and, in fact, it was the one used to first observe $b \to u$. The CLEO data is shown in Fig.~\ref{CLEO_Vub}. CLEO \cite{CLEO_end} and BaBar \cite{BaBar_end} report values of
\begin{eqnarray}
|V_{ub}|&=&(4.08\pm 0.34\pm 0.44 \pm 0.16 \pm 0.24)\times
10^{-3},   \\\nonumber
|V_{ub}|&=&(4.43\pm 0.29\pm 0.50 \pm 0.25 \pm 0.35)\times
10^{-3},
\end{eqnarray}
respectively, where the errors are experimental, $b\to u$, OPE and $b\to s\gamma$. Belle also presented a preliminary value \cite{Belle_end}.

There are additional theoretical errors that, however, have not been assigned by the experiments.
Bauer, Luke and Mannel point out that there is an additional uncertainty due to subleading twist contributions, basically higher order terms not accounted for using $b\to s\gamma$ for $f(k^+)$ \cite{BLM}. Their estimate of the additional error caused by these terms is $\sim$15\%. Furthermore, there is no assessment of the error due to weak annihilation which could be as large as 30\%.

Belle has used two other techniques to measure $|V_{ub}|$. One uses
$B\to D^{(*)}\ell\nu$ as a tag and the other uses neutrino reconstruction
\`a la exclusive semileptonic decays and then uses a sorting algorithm that they call ``annealing" to separate the event into a tag side and a $b\to u\ell\nu$ side.
To derive $|V_{ub}|$ in both cases they use $M_x~<~1.5$ GeV and for the second they add a $q^2~>$ 7 GeV$^2$ requirement \cite{Belle_end}. Their values are
\begin{eqnarray*}
|V_{ub}|&=&(5.00\pm 0.60\pm 0.23 \pm 0.05 \pm 0.39 \pm 0.36)\times
10^{-3},   \\\nonumber
|V_{ub}|&=&(3.96\pm 0.17\pm 0.44 \pm 0.34 \pm 0.26 \pm 0.29)\times
10^{-3},\\\nonumber
\end{eqnarray*}
respectively, where the errors are statistical, systematic, $b\to
c$, $b\to u$ and theoretical.

\subsubsection{\boldmath $|V_{ub}|$, Best Value and Error}
\label{subsec:Vubsum}

\begin{figure}[htb]
\centerline{\epsfig{figure=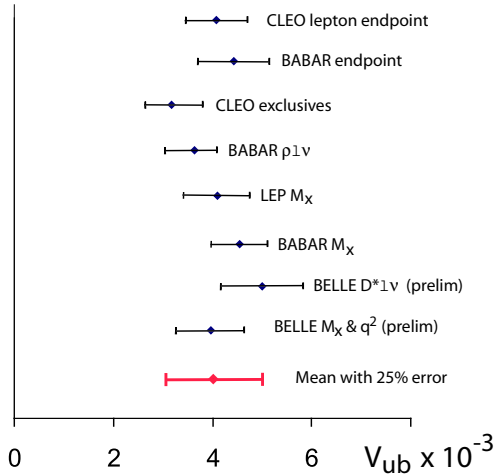,height=2.5in}}
\caption{Summary of $|V_{ub}|$ measurements.}
\label{Vub_sum}
\end{figure}

A summary of all the measurements with the quoted errors added in
quadrature is given in Fig.~\ref{Vub_sum}. We see that they are
nicely clustered with an r.m.s. of $\sim0.6\times 10^{-3}$.
However, it would be unwise to use this spread to assign an error
for several reasons. First of all, there are theoretical errors
that have not been included, and it seems that the more we learn about these decays the larger the errors become. Secondly, it is difficult to estimate
how early measurements affected the central values of the ones that followed.
Possibly it is safe to say that $|V_{ub}|=(4.0\pm 1.0)\times
10^{-3}.$  In the future there will be
more data from the B factories with excellent tagging, such as
demonstrated by the fully reconstructed tags from BaBar. Also,
unquenched Lattice calculations for exclusive final states should
become available in the large $q^2$ region.

I will now give a subjective evaluation of the current value of
$|V_{ub}|$.  Since we want to see if New Physics is present we need
to be conservative in assigning errors. The experimental statistical
and systematic errors first need to be evaluated. For
exclusive measurements I will use only quenched Lattice QCD
calculations to turn the measured rate into a value for
$|V_{ub}|$. These calculations have a 10\% intrinsic error to
which I add a 20\% quenching error to arrive at a 22\% total
error. For inclusive measurements HQE is used and I add in the
duality error, the error on $m_b$ and the weak annihilation error,
the sizes depending on the phase space region used by each
measurement.

The values for the different methods are listed in
Table~\ref{tab:Vub}. For exclusive final states I averaged CLEO
and BaBar $\rho\ell\nu$ and CLEO $\pi\ell\nu$, using only Lattice
QCD with a 22\% error.  For the lepton endpoint, the theoretical
error is taken as the quadrature of 10\%-$b\to u$, 3.7\%-$b\to
s\gamma$, 15\%-higher twist, 30\%-weak annihilation, 7\%-duality,
5\%-$m_b$. Measurements using only an $M_x$ $<$ 1.5 GeV are not
used due to the parton model singularity close to the cut and the
large resulting theoretical uncertainty. I do use the Belle
``annealing" result that has both the $M_x$$<$ 1.5 GeV cut and the
$q^2~>$ 7 GeV$^2$, where the theoretical errors are 7\%-duality,
7\%-weak annihilation and 10\%-$m_b$. The average value, that is
subjective but justifiable, is
\begin{equation}
|V_{ub}|=(3.90\pm 0.16_{exp}\pm 0.53_{thy}) \times 10^{-3}~~.
\end{equation}

\begin{table}
\caption{Evaluation of $|V_{ub}|$. The first error is experimental
and the second from theory.}
\label{tab:Vub}       
\begin{tabular}{lll}
\hline\noalign{\smallskip}
Method& Experiment &$|V_{ub}|$ $(\times 10^{-3})$ \\
\noalign{\smallskip}\hline\noalign{\smallskip}
Exclusives & CLEO \& BaBar &  $3.52\pm 0.27\pm 0.78$  \\
Lepton endpoint & CLEO \& BaBar & $4.28\pm 0.27\pm 1.44$\\
$M_x$ cut & LEP \& BaBar & Not used \cite{Defazio} \\
$M_x$ \& $q^2$ cuts & Belle & $3.96\pm 0.47\pm 0.56$\\
My Average & & $3.90\pm 0.16\pm 0.53$\\
\noalign{\smallskip}\hline
\end{tabular}
\end{table}

\section{\boldmath $B-\overline{B}$ and $D-\overline{D}$  Mixing }
\label{sec:Mixing}
\subsection{\boldmath $B_d$ and $B_s$ Mixing}
\label{subsec:BMixing}
Mixing in the $B_d$ system is very well measured \cite{PDG}.
The relationship between the measurement and the CKM matrix elements is given by
\begin{equation}
\label{eq:Bdmixing}
x\equiv{G_F^2\over 6\pi^2}B_Bf_B^2m_B\tau_B\left|V_{tb}^*V_{td}\right|^2
m_t^2F\left({m_t^2\over m_W^2}\right)\eta_{QCD},
\end{equation}
where $\eta_{QCD}$ is $\sim$0.8 and $F$ is a known function.  The
parameters $B_B$ and $f_B$ are currently determined only
theoretically. In principle $f_B$ can be measured, but its very
difficult. Finding ${\cal B}(B^+\to \tau^+\nu$) would measure the
product $f_B|V_{ub}|$. The best limit is $<4.1\times 10^{-4}$ from
BaBar \cite{Cartaro} that gives $f_B$ $<$ 390 MeV.

$B_s$ mixing is governed by a equation similar to
Eq. (\ref{eq:Bdmixing}), with all quantities referring now to the
$B_s$ and the important change that $V_{ts}$ appears rather than
$V_{td}$. This causes the oscillation rate to be rather high and
only a lower limit of $\Delta m_s > 14.4$ ps$^{-1}$ at 95\%
confidence level exists, compared with $\Delta m_d = 0.502\pm
0.006$ ps$^{-1}$ for $B_d$ \cite{Bsmixing}.

When $B_s$ mixing is measured we will learn about
\[
|V_{td}|^2/V_{ts}|^2=\left[(1-\rho)^2+\eta^2\right]\propto
f_{B_s}B_{B_s}^2/ f_{B_d}B_{B_d}^2 \equiv \zeta^2,
\]
which gives a circle in the $\rho-\eta$ plane centered at (1,0).
The theoretical ratio  on the right hand side is far easier to
calculate than the individual terms. Lattice QCD provides the best
values for $\zeta$. Wittig's summary gives $\zeta = 1.15 \pm
0.05^{+0.12}_{-0.00}$ \cite{Wittig}, while a recent partially
quenched calculation gives $1.14 \pm 0.03^{+0.13}_{-0.02}$
\cite{Aoki}. For fits used to find the values of $\rho$ and $\eta$
I will use $1.215\pm 0.030\pm 0.075$.

\begin{figure}[htb]
\vspace{-.4cm}
\centerline{\epsfig{figure=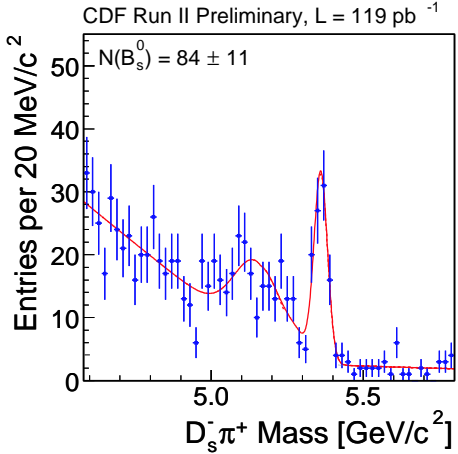,height=2.4in}}
\caption{The $D_s^+\pi^-$ candidate mass spectrum for
$D_s^+\to\phi\pi^+$ from CDF. The curve is a fit to the signal shapes for $\overline{B}_s\to D_s^+\pi^-$ (narrow peak) and $\overline{B}_s\to D_s^{*+}\pi^-$ (wide peak), where the $\gamma$ from the $D^{*+}_s$ decay is not observed. There are 84$\pm$11 events in the narrow peak.} \label{CDF_BstoDspi}
\end{figure}

CDF is making progress analyzing $B_s$. They have found the decay
$B_s\to D_s^+\pi^-$. Fig.~\ref{CDF_BstoDspi} shows their $B_s$
candidate mass spectrum \cite{DAuria}. The wider peak at lower
mass results from the decay $B_s\to D_s^{*+}\pi^-$. Here the decay
$D_s^+\to \phi\pi^+$ is used. CDF has measured the product of the
ratio of production ratios and two-body $B$ branching ratios as
\begin{equation}
{f_s\over f_d}\cdot {{{\cal{B}}(\overline{B}_s\to
D_s^+\pi^-)}\over {{\cal{B}}(\overline{B}^o\to D^+\pi^-)}}
=0.35\pm 0.05\pm 0.04 \pm 0.09~~.
\end{equation}
Accumulating these events is needed to measure $B_s$ mixing and
CDF is off to a good start.

\subsection{\boldmath $D^{\rm o}$ Mixing}
\label{subsec:Dmixing} 
\begin{figure}
\vspace{-.2cm}
\centerline{\epsfig{figure=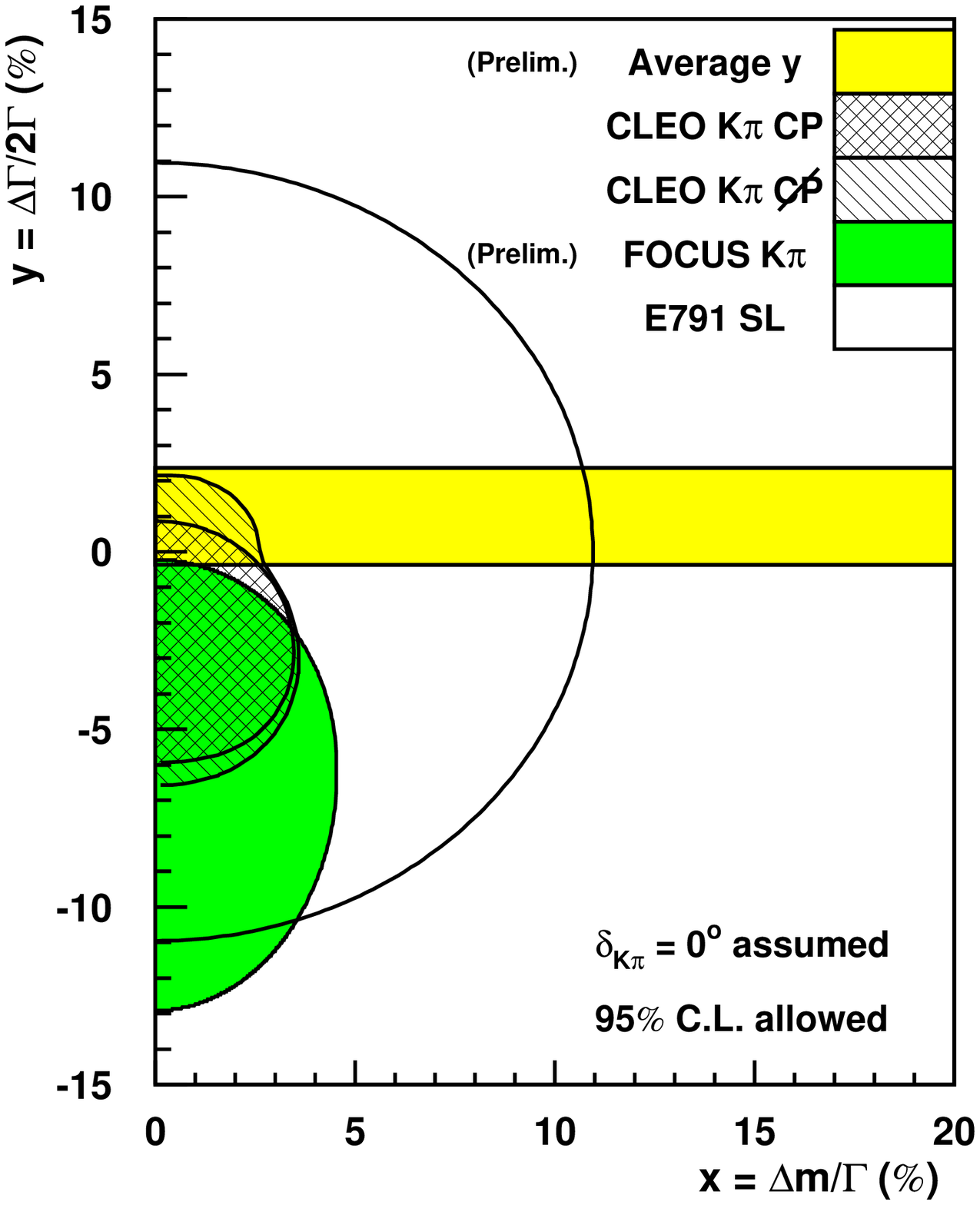,height=2.in}
\epsfig{figure=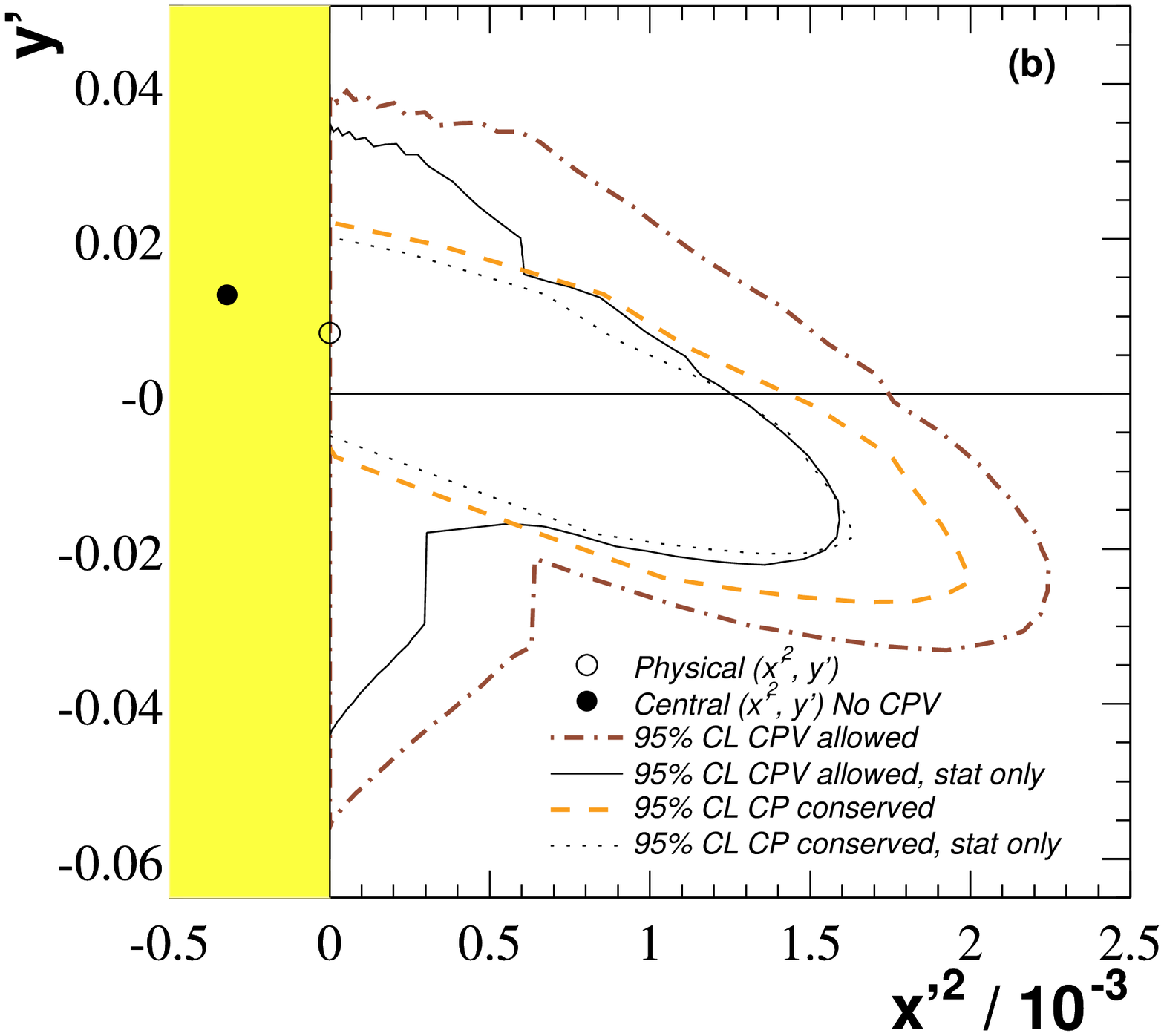,height=1.8in}}
\caption{Constraints on the size of $y=\Delta\Gamma/2\Gamma$ versus $x=\Delta m/\Gamma$
for both CP allowed and CP conserved scenarios from $D^{\rm o}$
measurements. (a) Shows several experiments, while (b)
is from BaBar only for $D^{\rm o}\to K^-\pi^+$. $x'$
and $y'$ refer to $x$ and $y$ allowing a rotation by an arbitrary
phase.}
\label{D0mix_all}       
\end{figure}

Several groups have searched for, but not
yet found mixing in the $D^{\rm o}-\overline{D}^{\rm o}$ system.
Measurements are made looking at the lifetime difference between
CP equal to +1 and -1 eigenstates given by the parameter
$y=\Delta\Gamma/2\Gamma$ and the mass difference between the two
eigenstates given by the parameter $x=\Delta M/\Gamma$. Standard
model predictions for these parameters are small, so this is a
good place to look for the effects of new physics
\cite{D0_mix_pred}. Alas, no evidence of such effects has been
uncovered. Upper limits from several experiments are shown in
Fig.~\ref{D0mix_all} \cite{D0_mixing}. Analyses are done both not allowing CP violation in this system and also by permitting it.

\section{Searches for New Physics: CKM Fits and Rare Decays}
\label{sec:NP}
\subsection{CKM Fits}
\label{subsec:Fits} There are many ways of looking for New Physics
\cite{Stone-NP} . One interesting way to is use different kinds of
measurements to determine the values of $\rho$ and $\eta$
\cite{Peskin}. We could in principle find  $\rho$ and $\eta$ using
only the magnitude measurements $|V_{ub}|$ and $B$ mixing and see
if different CP violating measurements give consistent values. We
should separately test CP violation in $K^{\rm o}_L$ decays, $B_d$
decays (i.e. using $B^{\rm o}\to J/\psi K_s$ and $B^{\rm o}\to
\rho \pi$)and $B_s$ decays (i.e. using $B_s\to D_s K$ and $B_s\to
J/\psi\eta$). Unfortunately we do not have these measurements at
our disposal yet, though there are future dedicated hadron
collider experiments, BTeV \cite{BTeV} and LHCb \cite{LHCb}, that
should provide them.

\begin{figure}[htb]
\vspace{-.2cm}
\centerline{\epsfig{figure=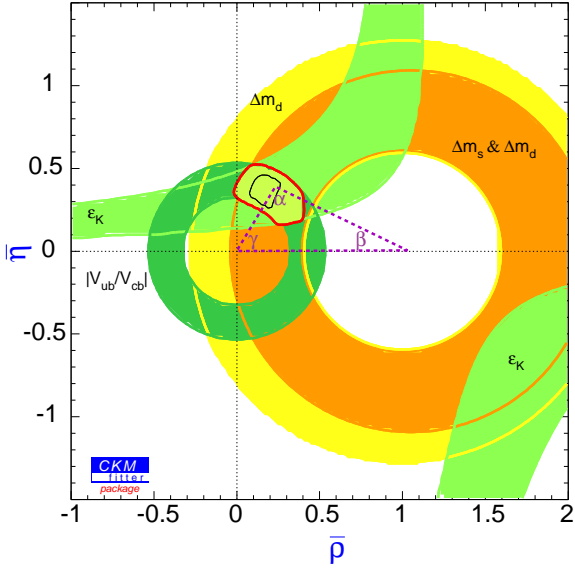,height=3.0in}}
\caption{Constraints on $\bar{\rho}$ and $\bar{\eta}$ using the values of the
variables given in this talk for $|V_{ub}|, |V_{cb}|$, the upper
limit on $B_s$ mixing and $B_d$ mixing, and $\epsilon_K$ from the
PDG \cite{PDG}. The outer circle is at 95\% c. l. and the inner
one at 32\%. (Note $\bar{\rho}=\rho(1-\lambda^2/2)$ and
$\bar{\eta}=\eta(1-\lambda^2/2)$.)} \label{rhoeta_ss1}
\end{figure}

Several groups have performed fits to the CKM parameters and the
methods are controversial. One set of groups, termed ``Bayesians"
treats the theoretical uncertainties as being Gaussian distributed
\cite{Bayesians}. Other groups eschew this prescription. They
object because of the difficulties in assigning errors to models
that have embedded assumptions. I use here the ``Rfit" method.
Theoretical errors are dealt with by restricting the theory
variable to a 95\% confidence interval with no preferred central
value \cite{Freq}.\footnote{This group has kindly made their
program available for use by others; see \cite{ckmfitter}.} Fits
for $\rho$ and $\eta$ using $|V_{ub}|$, $B_d$ mixing, the upper
limit on $B_s$ mixing and CP violation in the kaon system
$\epsilon_K$ give the result shown in Fig.~\ref{rhoeta_ss1}. This
is somewhat logically inconsistent since I am using $\epsilon_K$,
but this is what has been traditionally done.  We then can compare
with $\sin 2\beta$ \cite{Hitoshi}. Treatment of the theoretical
errors is particularly important here, as the theoretical errors
are larger than the experimental ones for most of the input
variables. Other non-Bayesian techniques give similar results as
Rfit \cite{Eigen}; in particular, Dubois-Felsmann \etal , treat
$B_s$ mixing in a more conservative manner and find a somewhat
larger region of acceptable values toward negative $\rho$.

\subsection{Rare $b$ Decays}
\label{subsec:rareb}
\subsubsection{Rare Electromagnetic $b$ Decays}

\begin{figure}[htb]
\vspace{-.6cm}
\centerline{\epsfig{figure=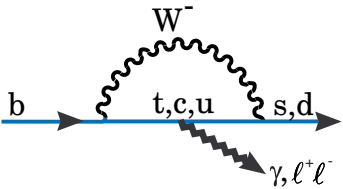,height=1.1in}}
\caption{Basic diagram for rare $b$ decay.} \label{loop_dia_rare}
\end{figure}

Rare $b$ decays are an excellent place to find new
physics. Fig.~\ref{loop_dia_rare} shows the basic structure of
these processes. Heretofore unknown fermion-like objects can
replace the quarks or new gauge-like objects can replace the
$W^-$.

\begin{figure}[htb]
\centerline{\epsfig{figure=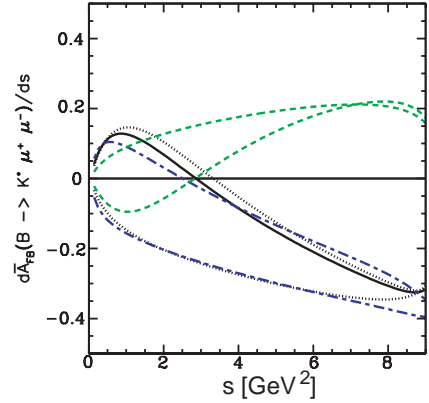,height=2.1in}}
\caption{Predicted dilepton asymmetry as a function of dilepton
mass squared (s) for the Standard Model (solid line) and different
SUSY models \cite{Alipol}.} \label{Ksmumu_asy_ss}
\end{figure}

Exclusive rare processes such as $B\to\rho\gamma$ and $B\to
K^*\ell^+\ell^-$ are important as well as inclusive processes. For
example Ali \etal ~show that SUSY can change the shape of the
polarization in $K^*\ell^+\ell^-$ (see Fig~\ref{Ksmumu_asy_ss})
\cite{Alipol}.

\begin{figure}[htb]
\centerline{\epsfig{figure=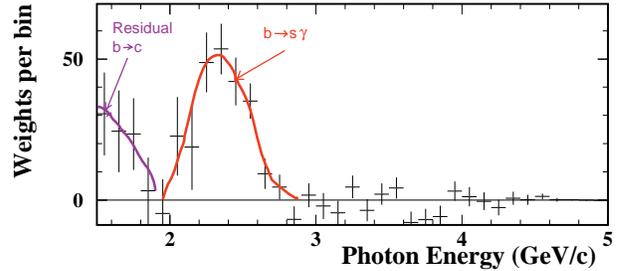,height=1.4in}}
\caption{Continuum subtracted photon momentum distribution for $b\to s\gamma$ from CLEO.} \label{btosg}
\end{figure}

The inclusive rate for $b\to s\gamma$ is important in its own
right and the shape of photon momentum spectra is used to get
information on the Fermi momentum distribution $f(k^+)$ of the $b$
quark in the $B$ meson, knowledge of which is needed for finding
CKM matrix elements. CLEO still has the best determination
\cite{CLEO_bsg}.
\begin{equation}
{\cal B}(b\to s\gamma)=(3.21 \pm 0.43\pm 0.27^{+0.18}_{-0.10})\times 10^{-4}~.
\end{equation}
The continuum subtracted photon momentum distribution is shown in
Fig.~\ref{btosg}.

Other measurements have been made by ALEPH, Belle and BaBar
\cite{Others_bsg}. The world average is
\begin{equation}
{\cal B}(b\to s\gamma)=(3.40\pm 0.39)\times 10^{-4}~.
\end{equation}

In lowest order the Hamiltonian can be written as
\begin{eqnarray*}
\cal{H}&=&{4G_F\over
\sqrt{2}}\left(V_{tb}V_{ts}^*\right)\left[c_7(m_b)O_7
+c_8(m_b)O_8\right], \\
O_7&=&{e\over 16\pi^2}
m_b\bar{s}_L\sigma_{\mu\nu}b_RF^{\mu\nu},~
O_8={1\over 4\pi} m_b\bar{s}_L\sigma_{\mu\nu}b_RG^{\mu\nu}~. \\\nonumber
\end{eqnarray*}
The resulting decay rate in the Standard Model is
\begin{equation}
\Gamma(b\to s\gamma)={G_F^2\alpha m_b^5 \over
32\pi^4}\left|c_7\right|^2 \left|V_{tb}V_{ts}^*\right|^2~.
\end{equation}

Theorists then take $|V_{ts}|=|V_{cb}|$ and calculate the next to
leading order (NLO) corrections. The full set of NNLO corrections
has not yet been fully calculated. We are left with two
predictions $(3.32\pm 0.30)\times 10^{-4}$ and $(3.70\pm
0.30)\times 10^{-4}$ (see Greub \cite{Greub}), which I average to
get a final SM theoretical prediction of $(3.5\pm 0.5)\times
10^{-4}$, completely consistent with the data.

\begin{figure}[htb]
\vspace{-.6cm}
\centerline{\epsfig{figure=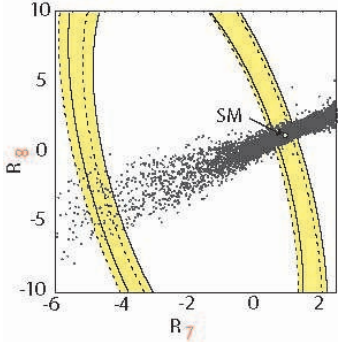,height=1.8in}}
\caption{Predictions of SM and MFV type SUSY models versus $R_7$ and $R_8$ (see text).
The constraint from $b\to s\gamma$ is shown by the bands.} \label{btosg_np}
\end{figure}

Many non-SM models are ruled out by this comparison. For example,
Ali \etal ~define the parameters
$R_i=(c_i^{SM}+c_i^{NP})/c_i^{SM}$, for $i=7,8$.
Fig.~\ref{btosg_np} shows the contours allowed by the data along
with the predictions of the SM and Supersymmetric Models with
Minimal Flavor Violation \cite{Ali_npsg}.

Belle first observed the dilepton decays in the $K\mu^+\mu^-$
final state \cite{Bellemumu}. Evidence for $K^*\mu^+\mu^-$ at the
3$\sigma$ level was shown at this conference by BaBar \cite{Ryd};
recently Belle also has shown a signal in this mode
\cite{Belle-Ksll}. Belle has also measured inclusive
$X_s\ell^+\ell^-$ \cite{Belleincl}. The branching ratios given in
Table~\ref{tab:xll} are in agreement with SM predictions, but have
large errors due to small statistics. For example, Belle has 30
$K^{*{\rm o}}\ell^+\ell^-$ events in 140 fb$^{-1}$; clearly much
larger samples are needed to probe for new physics.

\begin{table}
\begin{center}
\caption{Branching ratios for rare dilepton decays ($\times
10^{-7}$). }
\label{tab:xll}       
\begin{tabular}{lll}
\hline\noalign{\smallskip}
Reaction & Belle & BaBar \\
\noalign{\smallskip}\hline\noalign{\smallskip}

$B\to K\ell^+\ell^-$ &  $4.8^{+1.0}_{-0.9}\pm$0.3$\pm$0.1
&$6.8^{+1.7}_{-1.5}\pm$0.4 \\
$B\to K^*\ell^+\ell^-$  & 11.5$^{+2.6}_{-2.4}\pm$0.8$\pm$0.2
&$14.0^{+5.7}_{-4.9}\pm$2.1\\
$B\to X_s\ell^+\ell^-$ & 61$\pm$$14^{+14}_{-11}$ & - \\
\noalign{\smallskip}\hline
\end{tabular}
\end{center}
\end{table}

\subsubsection{Rare Hadronic $b$ Decays}

First I will discuss some channels that do not proceed via loop
diagrams. Let us consider phase shifts. In $D$ decays the phase
shifts are known to be large. In $B$ decays CLEO \cite{CLEOD0pi0}
and Belle \cite{BelleD0pi0} observed $\overline{B}^{\rm o}\to
D^{\rm o}\pi^{\rm o}$ which allowed the determination of the phase
shift between the $\Delta I=3/2$ and $\Delta I=1/2$ amplitudes
using the other two legs of the isospin triangle found from
$\overline{B}^{\rm o}\to D^+\pi^-$ and $B^-\to D^{\rm o}\pi^-$.
The phase shift is found to be between 16.5$^{\circ}$ and
38.1$^{\circ}$ at 90\% c. l. BaBar has also measured these decays
\cite{BaBarD0pi0}.

\begin{figure}[htb]
\vspace{-.5cm}
\centerline{\epsfig{figure=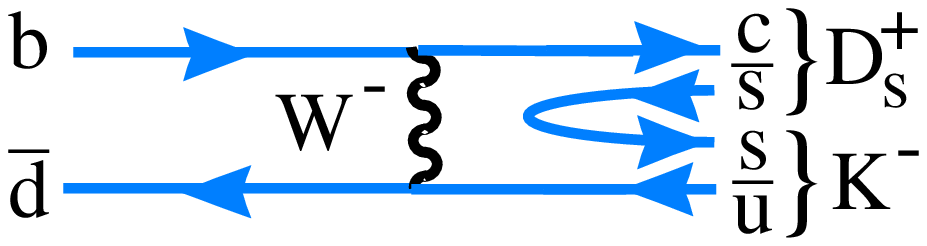,height=.7in}}
\caption{$W$ exchange diagram for $\overline{B^{\rm o}}\to D_s^+
K^-$.} \label{BtoDsK1}
\end{figure}

Belle \cite{Belle_DsK} and BaBar \cite{Babar_DsK} have also
observed the decay $\overline{B^{\rm o}}\to D_s^+ K^-$ at a level
of $4\times 10^{-5}$ which can occur via the $W$ exchange diagram
shown on Fig.~\ref{BtoDsK1} or could be a result of rescattering
from $D^+\pi^-$, for example. Phase shifts and rescattering go
hand-in-hand.

\begin{figure}[htb]
\vspace{-.6cm}
\centerline{\epsfig{figure=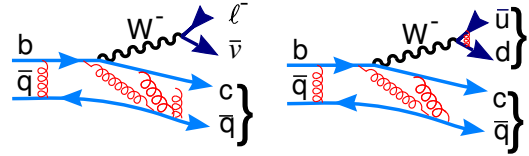,height=0.9in}}
\caption{Exclusive semileptonic decay and two-body
hadronic decay into a $D^{(*)}$ plus light hadron. Gluons connecting
the $\bar{u}$ or $d$ with the $c$ or $\bar{q}$ are ignored, hence
factorization.}
\label{factor1}
\end{figure}

We now consider the concept of ``factorization." This is a
fundamentally simple idea, that the amplitude in two-body hadronic
decays is a product of two hadronic currents similar to
semileptonic decays (see Fig.~\ref{factor1}) where there is one
hadronic current and one leptonic current \cite{Bjorken}. If
factorization is a valid concept then the we can compare the decay
widths for the hadronic two-body decay at $q^2$ equal to the
mass-squared of the light hadron ($h$) according to the formula
\begin{equation}
\Gamma\left(\overline{B}\to D^*h^-\right) =
6\pi^2a_1^2f_h^2|V_{ud}|^2{d\Gamma \over dq^2}
(\overline{B}\to D^*\ell^-\bar{\nu})\left|_{q^2=m_h^2}\right.,
\end{equation}
where $f_h$ is the decay constant for the light hadron and $a_1$
is a theoretical parameter. Early tests assumed that $a_1$ should
be unity \cite{Borto}. In BBNS, a modern theory of factorization,
$a_1=1.05$ is a precisely calculated value \cite{BBNS}.

Table~\ref{tab:fact} compares available data on three hadronic
modes with the semileptonic data from CLEO \cite{Ecklund}.
Unfortunately, other groups have not published their values of
$d\Gamma/dq^2$.

\begin{table}
\caption{Factorization Tests. }
\label{tab:fact}       
\begin{tabular}{lccc}
\hline\noalign{\smallskip}
Mode & $\overline{B}^{\rm o}\to D^{*+}\pi^-$ &
$\overline{B}^{\rm o}\to D^{*+}\rho^-$&$\overline{B}^{\rm o}\to D^{*+}a_1^-$ \\
\noalign{\smallskip}\hline\noalign{\smallskip}
${\cal{B}}$(\%) & 0.28$\pm$0.02 &0.68$\pm$0.09 & 1.30$\pm$0.27 \\
$d\Gamma/dq^2$ &
2.12 & 2.36 &2.76\\
{\tiny (ns$^{-1}$GeV$^{-2}$)} &{\scriptsize $\pm 0.22\pm 0.21$}  & {\scriptsize $\pm 0.22\pm 0.21$} & {\scriptsize $\pm 0.20\pm 0.22$} \\
$f_h$ (MeV) & 131.7 & 215 & 205 \\
$a_1$ &0.93$\pm$0.07 & 0.85$\pm$0.07 &0.98$\pm$0.11 \\
\noalign{\smallskip}\hline
\end{tabular}
\end{table}

The values of $a_1$ average about 14\% below the BBNS predicted
value. However, the CLEO value for ${\cal{B}}(\overline{B}^{\rm
o}\to D^*\ell\bar{\nu}$) is 16\% higher than the world average
\cite{HFAG}, implying that $a_1$ could be raised by 8\% if other
measurements of $d\Gamma/dq^2$ were available. Thus I conclude
that the BBNS predicted value of $a_1$ is likely consistent with
the data.

\begin{figure}[hbt]
\vspace{-.7cm}
\centerline{\epsfig{figure=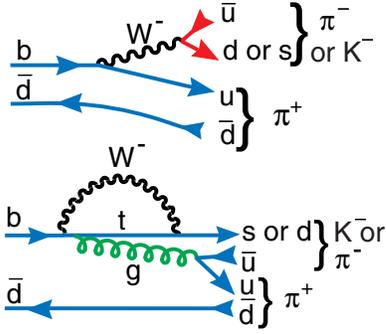,height=1.8in}}
\caption{Diagrams leading to charged $K\pi$ final states. (top)
Decays via $b\to u$ and (bottom) Penguin decays.} \label{Kpi-pipi}
\end{figure}

Now we turn to two-body rare decays into light hadrons. The
leading diagrams are shown in Fig.~\ref{Kpi-pipi}. One is a simple
spectator decay via $b\to u$ and the other is a loop
decay. Since both diagrams can lead to the same final states,
interference can occur.
\begin{figure*}[bht]
\centerline{\epsfig{figure=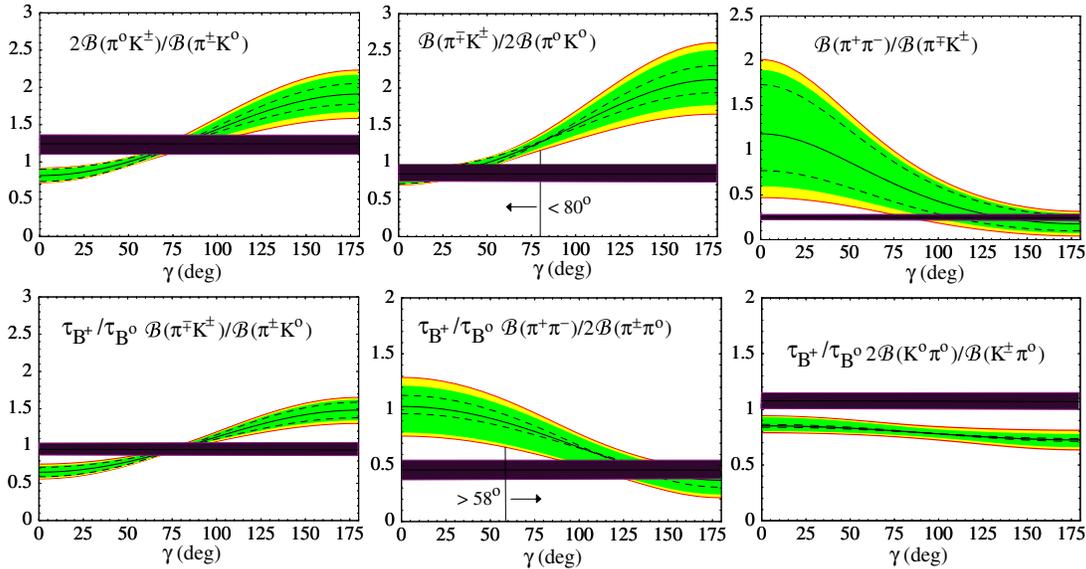,height=3in}}
\caption{Predictions from BBNS shown as curved bands and the world
average data shown as horizontal bands (central value $\pm
1\sigma$) as a function of $\gamma$. The vertical bands on the
center two plots indicate the values of $\gamma$ where the
measurements differ by 2$\sigma$ from the edges of the theory
bands.} \label{fig:BBNS}
\end{figure*}

These decays have been studied by several authors in a model dependent manner
\cite{Rosner-rare} and in the context of a QCD
factorization theory by BBNS \cite{BBNS}, who take the amplitude
involving both the $b$ and spectator quarks plus a part from the
virtual $W^-$ with corrections parameterized in a series
$\propto\sum(\Lambda_{QCD}/m_b)^n$. They compute the amplitudes and
the interferences for both the Tree and Penguin diagrams.
Averaged experimental branching ratios are given in Table~\ref{tab:twob} \cite{twob}.

\begin{table}
\caption{Branching ratios for $K\pi$ and $\pi\pi$ modes ($\times
10^{-7}$).}
\label{tab:twob}       
\begin{tabular}{lcccc}
\hline\noalign{\smallskip}
Mode &CLEO \cite{Gao_pp}& BaBar \cite{Ocariz} & Belle \cite{Piil}  & Average \\
\noalign{\smallskip}\hline\noalign{\smallskip}

$\pi^+\pi^-$ & $45^{+14+5}_{-12-4}$
 & 47$\pm$6$\pm$2 &44$\pm$6$\pm$3
 & 45.5$\pm$4.4\\
$\pi^+\pi^{\rm o}$ & $46^{+18+6}_{-16-7}$
 & $55^{+10}_{-~9}\pm$6 &53$\pm$13$\pm$5
 & 53$\pm$8\\
$K^{\pm}\pi^{\mp}$ & $188^{+23+12}_{-21-~9}$
 & 179$\pm$9$\pm$7 &185$\pm$10$\pm7$ & 183$\pm$7\\
$K^{+}\pi^{o}$ & $129^{+24+12}_{-22-11}$
 & $128^{+12}_{-11}\pm$10 &128$\pm$14$^{+14}_{-10}$ & 128$\pm$11\\
$K^{\rm o}\pi^{-}$ & $188^{+37+21}_{-33-18}$
 & 200$\pm$16$\pm$10 &220$\pm$19$\pm$11 & 206$\pm$13\\
$K^{\rm o}\pi^{o}$ & $128^{+40+17}_{-33-14}$
 & 104$\pm$15$\pm$8 &126$\pm$24$\pm$14 & 112$\pm$14\\
$\pi^{\rm o}\pi^{\rm o}$ &  $<$47  & 21$\pm$6$\pm$3 &
17$\pm$6$\pm$3 & 19$\pm$5 \\
\noalign{\smallskip}\hline
\end{tabular}
\end{table}

The interference between the Tree ($\propto V_{ub}$) and the
Penguin diagrams introduces the phase $\gamma$ into the prediction
of the decay rates. Discussing ratios rather than absolute rates
reduces the errors. Some BBNS predictions are compared with the
data from Table~\ref{tab:twob} in Fig.~\ref{fig:BBNS}. We see that
two of these ratios place restrictions of
$80^{\circ}~>~\gamma~>~58^{\circ}$, using 2$\sigma$ as limiting
the difference between the theory and data.

Certain other ratios present problems for this theory, however.
The $K^{\rm o}\pi^{\rm o}/K^+\pi^{\rm o}$ rate shown in the lower
right hand corner, is relatively insensitive to $\gamma$, yet
differs by more than 2$\sigma$ from the prediction for
$\gamma>58^{\circ}$. BaBar and Belle recently observed $B^{\rm
o}\to \pi^{\rm o}\pi^{\rm o}$ \cite{Fry}. The prediction for
$\tau_{B^+}/\tau_{B^{\rm o}}{\cal{B}}(\pi^{\rm o}\pi^{\rm o})/
{\cal{B}}(\pi^{\pm}\pi^{\rm o})$ is $<0.12$ for $\gamma <
80^{\circ}$ and $<0.27$ for all $\gamma$. The measured ratio is
0.42$\pm$0.11, presenting another contradiction, although the
$\pi^{\rm o}\pi^{\rm o}$ is particularly difficult to predict
because it is a low branching ratio color suppressed mode
\cite{Buchalla}. Since BBNS is a true theory, i. e. it makes
predictions based on general principles and prescribes a
convergent series approximation, then if future data do indeed
continue to show inconsistencies with this theory the reasons for
the theory breakdown must be understood. One possibility is that
there is new physics present. A recent paper that approaches these
decays in a different manner presents some evidence for new
physics \cite{Buras}.

Ignoring these caveats, the allowed range
$58^{\circ}>\gamma>80^{\circ}$ is in excellent agreement with the
allowed range of $\rho$ versus $\eta$ found by using the Rfit
method, shown in Fig.~\ref{fig:BBNS}. In fact this range is almost
identical with the 32\% confidence level allowed region outlined
in the figure. Chua \etal~ have an alternative model that
explicitly includes final state rescattering \cite{Hou}. Using a
similar set of reactions as BBNS, they find that $\gamma$ is in
the range of 90$^{\circ}$-100$^{\circ}$, which is barely
consistent with the values of $\rho$ and $\eta$ found by using
Rfit, and more consistent with the range found by Dubois-Felsmann
\cite{Eigen}.

\section{Revelations about QCD}
Since QCD is so important for extracting quark parameters, it is
interesting to see how well it's doing in other areas. These
include new narrow excited $D_s$ states, doubly charmed baryons
\cite{Selex}, measurements of the $\eta(2C)$ mass \cite{etacp}, D
states in the Upsilon system \cite{tomasz}  and  $D^{**}$ states
in $B$ decays \cite{BelleDss}. Unfortunately, space allows only a discussion of one of these interesting topics.

\subsection{The Narrow Excited $D_s$ States}
Excited $D_s$ mesons have unit angular momentum between the $c$
and $\bar{s}$ quarks. Different couplings to the quark spin give
predicted spin-parity, $J^p$, of: $0^+$, $1^+$, $1^+$ and $2^+$.
One $1^+$ and the $2^+$ have previously been seen. These decay into
$D^{(*)}K$, and are relatively narrow. Other states were also
predicted by most potential models to be above $D^{(*)}K$
threshold and have large $\sim$200 MeV widths \cite{poten-models}.

\begin{figure}[htb]
\centerline{\epsfig{figure=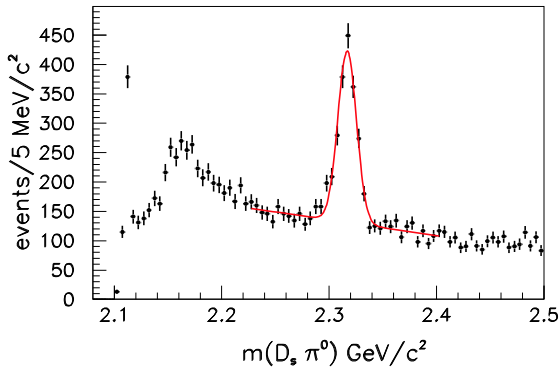,height=1.8in}}
\vspace{.1cm} \caption{The $D_s^+\pi^{\rm o}$ mass distribution
from BaBar. The very narrow peak near threshold is from the
isospin violating decay $D_s^{*+}\to \pi^{\rm o} D_s$.}
\label{Babar_Dspi0}
\end{figure}

BaBar observes a ``narrow" peak in the $D_s^+\pi^{\rm o}$ mass
distribution, shown in Fig.~\ref{Babar_Dspi0} \cite{BaBarDss}. The
mass is 2316.8$\pm$0.4$\pm$3.0 MeV, where the first error is
statistical and the second systematic. The width is consistent
with the mass resolution of $\sim$9 MeV (r.m.s.). The mass is
lighter than most potential model predictions and is 40 MeV below
$DK$ threshold.

Most potential models predicted the mass of this state to be above
$DK$ threshold. Those authors who did predict lower masses, did
not point out that the state would be narrow (see for example
\cite{Riz}, \cite{Deand}). After the announcement of this
discovery several possible explanations appeared, some of which
were quite exotic.  Barnes \etal ~argued the possibility of a $DK$
molecule \cite{Barnes}, while Szczepaniak argued for a $D\pi$ atom
\cite{Szcz}. Van Beveren and Rupp use a unitarized meson model to
explain the narrow mass as a kind of threshold effect \cite{VB}.
Cahn and Jackson formulate an acknowledgely poor explanation using
non-relavistic vector and scalar exchange forces~\cite{Cahn}.
Several authors propose a four-quark explanation \cite{4quark}.
Bardeen, Eichten and Hill \cite{bardeeneichtenhill} explain
the 2317 MeV object as an ``ordinary" $c\bar{s}$ state, that is
narrow only because isospin is violated in the decay.\footnote{Isospin is violated because the $D_s^+$ and its excitations are I=0, while the pion is I=1.} The isospin
violating channel is the only way for this state to decay since
the mass is below $DK$ threshold \cite{Cho-wise}. They use HQET
plus chiral symmetry to predict ``parity doubling," where two
orthogonal linear combinations of mesons transform as
SU(3)$_L\times$SU(3)$_R$ and split into ($0^-,1^-$), ($0^+,1^+$)
doublets.  Assuming that the $D_{sJ}^*(2317)$ is the $0^+$ state
expected in the quark model, they predict that the mass splitting
between the remaining $1^+$ state and the $1^-$ should be the same
as the $0^+-0^-$ splitting (see also \cite{NRZ}).

\begin{figure}[htb]
\begin{center}
\vspace{-4mm}
\epsfig{file=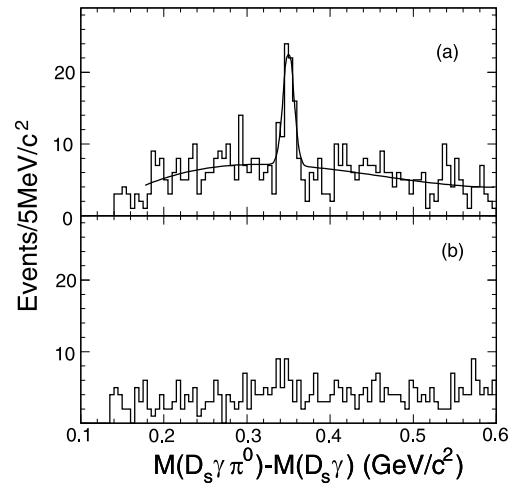,height=2.5in} \caption{The
$D_s^{*+}\pi^{\rm o}$ candidate mass distribution from CLEO shown
as the difference with respect to the $D_s^{*+}$ mass. (a)
$D_s^{*+}$ signal region; (b) $D_s^{*+}$ sideband region. }
\label{Stone-fig2}
\end{center}
\end{figure}

CLEO confirms the $D_s^+\pi^{\rm o}$ state seen by BaBar. They
find, that the measured width of the peak is $8.0^{+1.3}_{-1.2}$
MeV, somewhat wider than the detector resolution of $6.0\pm0.3$
MeV \cite{CLEO_Dss}. More interestingly, they also show
unequivocal evidence of a second state decaying into
$D_s^{*+}\pi^{\rm o}$ at a mass near 2460 MeV (see
Fig.~\ref{Stone-fig2}). The measured width is $6.1\pm$1.0 MeV,
close to the detector resolution of 6.6$\pm$0.5 MeV. There are
55$\pm$10 events in the peak. Although the BaBar data also showed
an excess of events in this mass region, the conclusion reached in
Ref.~\cite{BaBarDss} was that further study was needed to resolve
whether the peak received contributions from a new state or was
entirely due to a reflection of the $D_{sJ}^*(2317)$. I will
consider the questions of reflections and backgrounds next.

No known source has been identified that can create these narrow
peaks, other than new resonances. However, in both cases the mass
differences between the excited states and the $D_s$ or $D_s^*$
are about 350 MeV. This makes it easy for one state to reflect
into another. The real question is how much of each observed peak
is a reflection of the the other state. CLEO has two methods. In
their first method they perform a Monte Carlo simulation of
feed-down and feed-up. It turns out that the efficiency for the
higher mass peak to appear near the lower mass peak by simply
ignoring the photon from the $D_s^*$ decay is quite high,
(84$\pm$4$\pm$10)\%, but the predicted width of the peak is 14.9
MeV, rather larger than the observed width. By comparison, the
feed-up is rather small, (9.0$\pm$0.7$\pm$1.5)\% of the real signal in the lower mass peak, because a random
photon must be picked up to form a $D_s^*$. The smallness of this
feed-up is confirmed by small peaking in the $D_s^*$ sidebands, Fig.~\ref{Stone-fig2}(b),
which represents about 20\% of the total number of events in the signal peak.

In their second method CLEO fits the peak near 2317 MeV to two
Gaussians, one from the feed-down and one for the narrow state;
for the 2460 MeV state they perform a $D^{*+}$ sideband
subtraction before they fit the signal peak. In the 2317 MeV
region, their fit determines the signal to be at a mean mass
difference of 350.0$\pm$1.2$\pm$1.0 MeV with a width of
5.9$\pm$1.2 MeV and another wider Gaussian at 344.9$\pm$6.1 MeV
with a width of 16.5$\pm$6.3 MeV, characteristic of the feed-down
background. The systematic error on the mass of 1.0 MeV is smaller
than the 3.0 MeV assigned by BaBar because CLEO has removed the
effects of the feed-down background on the mass determination. The
sideband subtracted fit in the 2460 MeV region gives a mass
difference of 351.2$\pm$1.7$\pm$1.0 MeV.

\begin{figure}
\centerline{\epsfig{figure=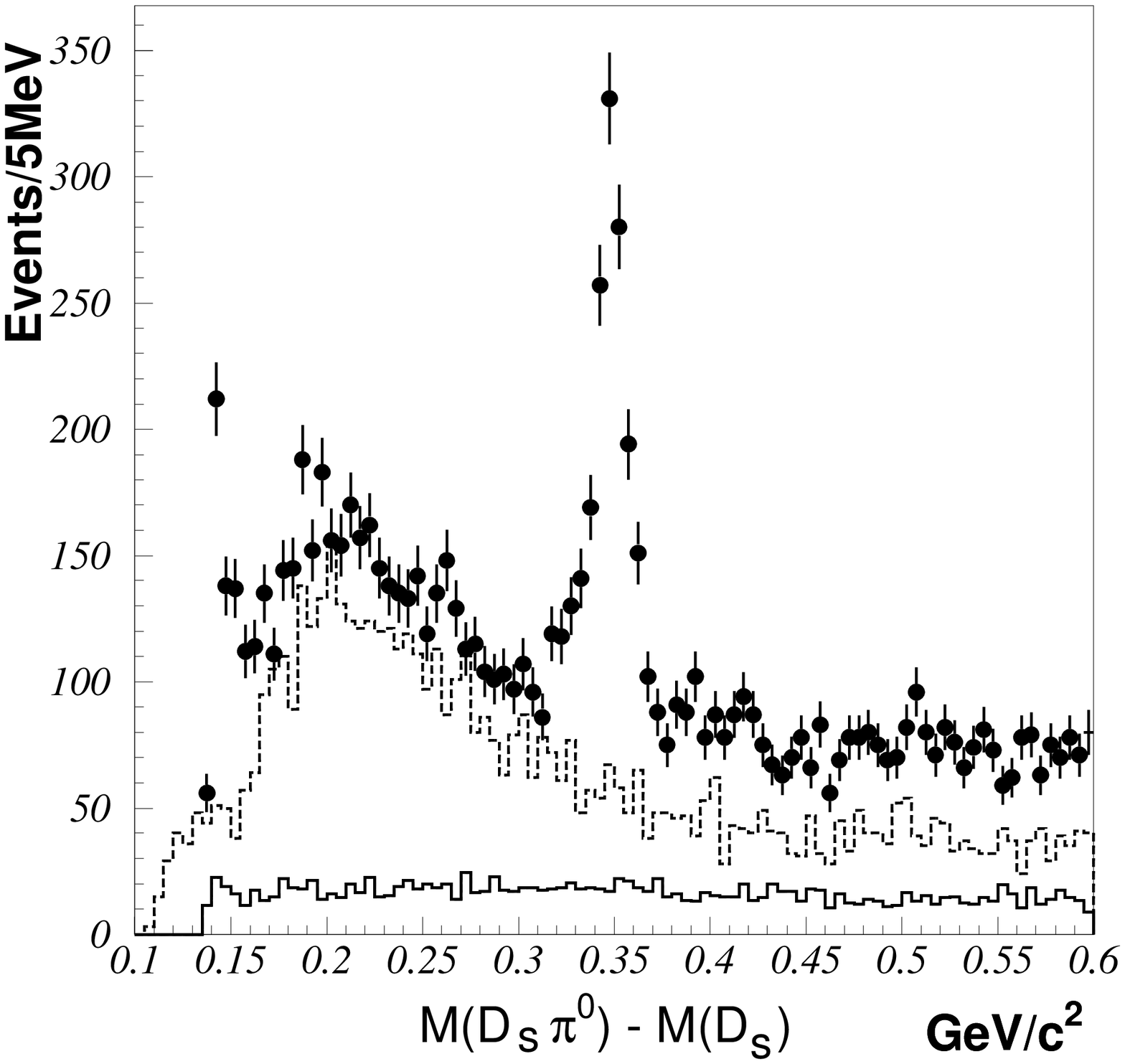,height=1.7in}
\epsfig{figure=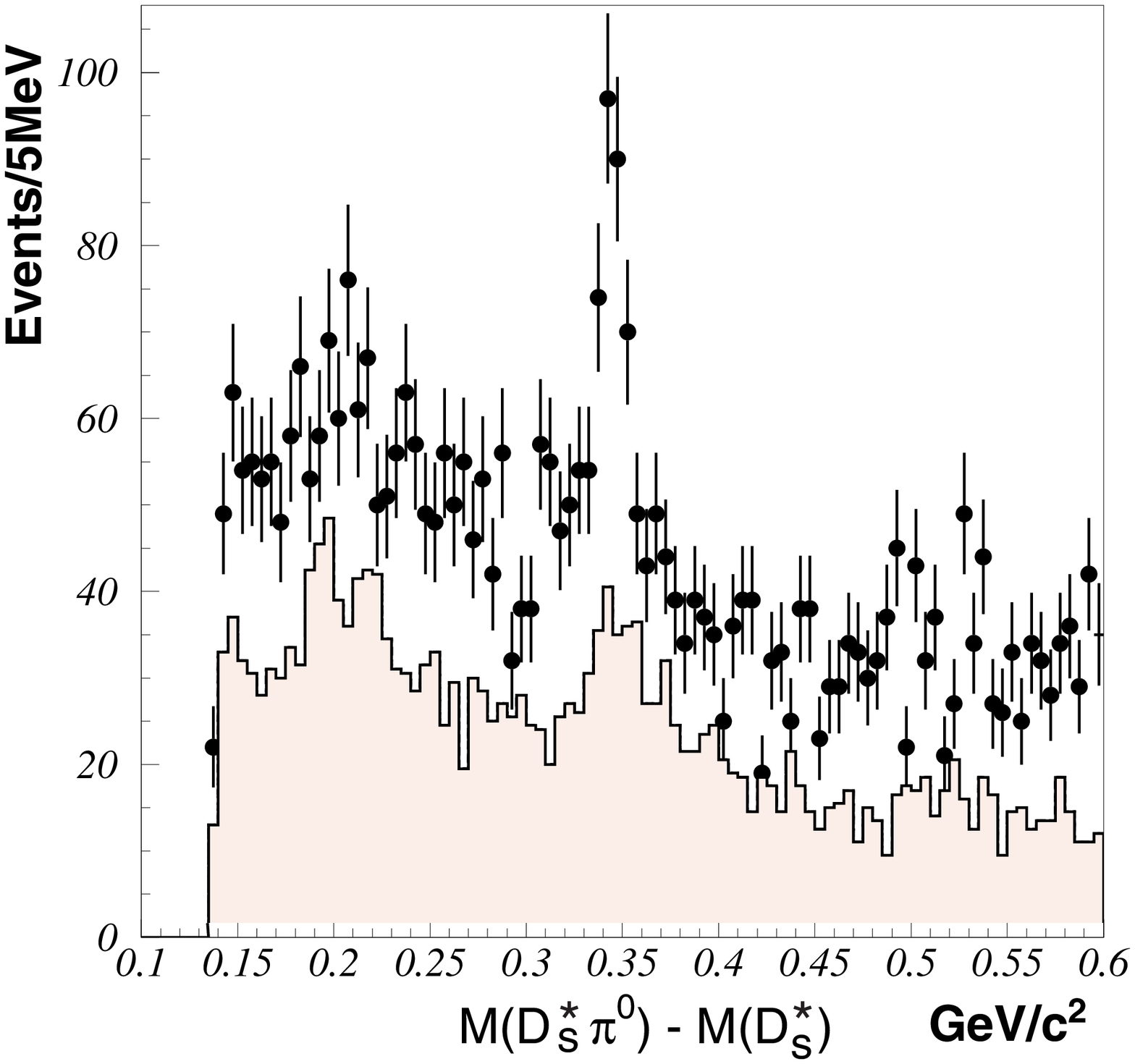,height=1.7in}}
\caption{Mass difference distributions from Belle. The histograms
are (left) $D_s^+$ sidebands (dark dotted line) and $\pi^{\rm o}$
sidebands
 (light dotted line) and (right) $D_s^{*+}$ sidebands.}
\label{Belle_dm}       
\end{figure}

Belle confirms both peaks in continuum $e^+e^-$ collisions
\cite{Seuster}. The Belle data are shown in Fig.~\ref{Belle_dm}.
There is clear peaking in the $D_s^{*+}$ sideband region showing
the level of feed-up. Furthermore, BaBar also now confirms the
existence of the $D_s^{*+}\pi^{\rm o}$ state \cite{Porter}.

\begin{figure}[htb]
\vspace{-6mm}
\begin{center}
\epsfig{file=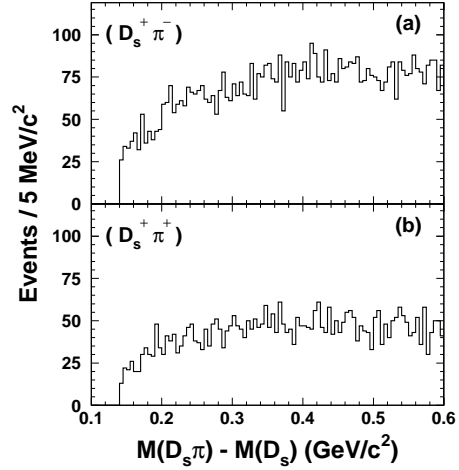,height=2.9in}
\vspace{-4mm}\caption{(Mass difference distributions from CLEO
for (a)  $D_s^{\pm}\pi^{\mp}$ and (b)$D_s^{\pm}\pi^{\pm}$. }
\label{Stone-fig4}
\end{center}
\end{figure}

Thus, there is no question about the existence of such states; we
do need still to investigate what they are and what they tell us
about QCD. CDF \cite{Majorie} and CLEO have looked for similar
neutrally charged states in $D_s^{\pm}\pi^{\mp}$ and doubly
charged states in $D_s^{\pm}\pi^{\pm}$. No signals were found. The
CLEO data is shown in Fig.~\ref{Stone-fig4}; upper limits over the
shown mass difference range are better than a factor of ten lower
than the observed $D_s^+\pi^{\rm o}$ signal \cite{Urheim}. CLEO
also finds upper limits on many other decay channels of both
states. The lack of any isospin partner states casts doubt on any
molecular explanation.

\begin{figure}[htb]
\vspace{-5mm}
\begin{center}
\epsfig{file=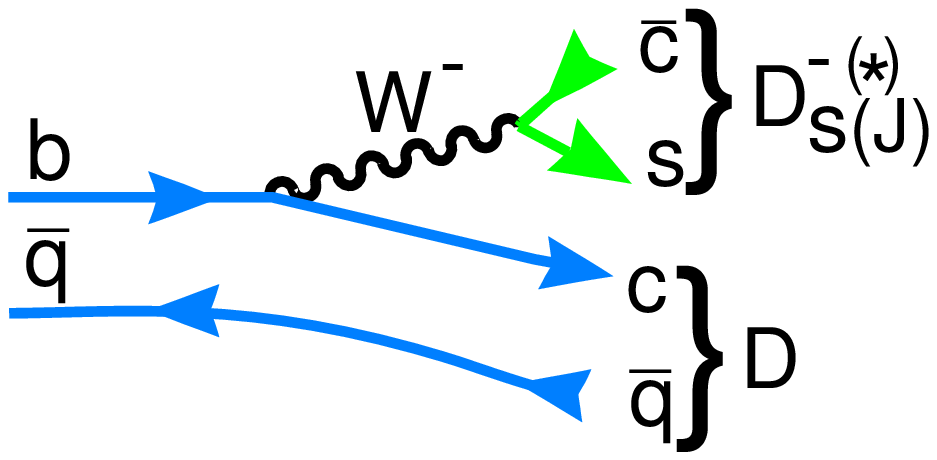,height=1in} \caption{Feynman diagram for
$B\to D D_{sJ}^{(*)-}$.} \label{Diag-DDs}
\end{center}
\end{figure}

\begin{figure}[htb]
\vspace{-5mm}
\begin{center}
\epsfig{file=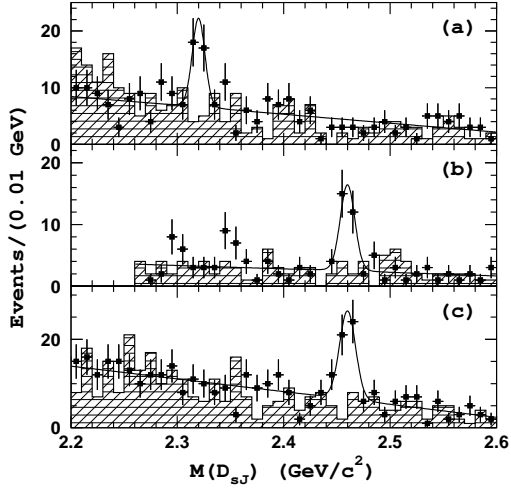,height=2.7in} \caption{Invariant
masses for $D_{sJ}^{(*)-}$ candidates produced in the reaction
$B\to D D_{sJ}^{(*)-}$. The open regions are signal, the cross
hatched regions combinations of $D_{sJ}^{(*)}$ mass and $\Delta E$
sidebands for (a) $D_s\pi^{\rm o}$, (b) $D_s^*\pi^{\rm o}$ and (c)
$D_s\gamma$. } \label{Belle_Dss_mass}
\end{center}
\end{figure}

These states should also be seen in $B$ decays. The modes $B\to
D^{(*)}D_{s}^{(*)-}$ have been observed long ago \cite{PDG}.
Lipkin, in fact, predicted that the $1^+$ states would be produced
in the reaction $B\to D D_{sJ}^{(*)-}$ \cite{Lipkin}. The diagram
for such processes is shown in Fig.~\ref{Diag-DDs}. Belle has
observed these reactions \cite{Belle_BtoDsj}.
Fig.~\ref{Belle_Dss_mass} shows the reconstructed $D_s^+\pi^{\rm
o}$, $D_s^{*+}\pi^{\rm o}$ and $D_s^+\gamma$ mass spectrum for
events whose mass and beam energy constraints are consistent with
the reaction $B\to D D_{sJ}^{(*)-}$.

The peak in Fig.~\ref{Belle_Dss_mass}(c) represents the first
observation of the radiative decay mode of the $D_{sJ}(2460)$. The
product branching ratios are given in Table~\ref{tab:BtoDDsj}.

\begin{table}
\caption{${\cal{B}}(B\to D D_{sJ}^{(*)})$ from Belle}
\label{tab:BtoDDsj}       
\begin{center}
\begin{tabular}{ccc}
$B$ mode & $D_{sJ}$ mode & ${\cal{B}}\times 10^{-4}$\\
\hline\noalign{\smallskip}
 $DD_{sJ}^{*-}(2317)$ &$\pi^{\rm o}D_s$ &$8.5^{+2.5}_{-1.9}\pm 2.6$ \\
 $DD_{sJ}^{-}(2460)$ &$\pi^{\rm o}D_s$ &$17.8^{+4.5}_{-3.9}\pm 5.3$ \\
 $DD_{sJ}^{-}(2460)$ &$\gamma D_s$ &$6.7^{+1.3}_{-1.2}\pm 2.0$ \\
\noalign{\smallskip}\hline
\end{tabular}
\end{center}
\end{table}

The relative width of the radiative to the isospin violating decay
is
\begin{equation}
{{\Gamma\left(D_{sJ}(2460\to D_s\gamma\right)} \over
{\Gamma\left(D_{sJ}(2460\to D_s\pi^{\rm o}\right)}} = 0.38\pm
0.11\pm 0.04~~.
\end{equation}
Another determination in the continuum by Belle, gives a somewhat
inconsistent value of $0.63\pm 0.15 \pm 0.15$ \cite{Belle_cont};
the average value is $0.44\pm 0.10$.

The branching ratio of $B\to D D_s^{(*)}+$ is $\sim$1\%
\cite{PDG}. Assuming that the decay modes shown in
Table~\ref{tab:BtoDDsj} are dominant, then the branching ratios to
these excited states are about a factor of 4-10 lower. Chen
and Li \cite{Chen-Li} and Cheng and Hou \cite{4quark} predict that
a four-quark state would have one order of magnitude lower
branching ratio. Datta and O'Donnell agree that factorization
predicts a similar rate for the excited states as the $D_s^+$ and
show that a molecular explanation is consistent with the data
\cite{Datta}. Fundamentally, the disagreement with factorization
arises out of assuming that the decay constant of these excited
states is the same as the that of the $D_s^+$; then taking either
a molecular or four-quark structure for the new states allows the
coupling to the virtual $W^-$ to be smaller and thus explains the
data.

\begin{figure}
\vspace{-.4cm}
\centerline{\epsfig{figure=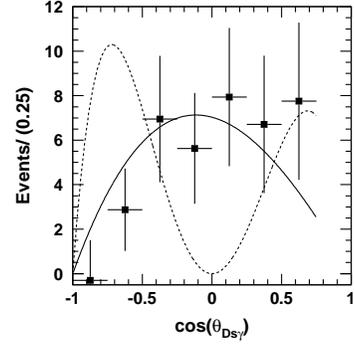,height=1.9in}}
\caption{Angular distribution in the reaction $B\to
DD_{sJ}^{-}(2460)$, $D_{sJ}\to\gamma  D_s^-$ of the $D_s$ in the
$D_{sJ}$ rest-frame with respect to the $D_{sJ}$ direction in the
$B$ rest-frame. The points with error bars are the data, the solid (dashed)
curve the expectation for a $1^+$ ($2^+$) $D_{sJ}$ state.}
\label{Belle_Dss_hel}       
\end{figure}

The spin-parity, $J^P$, of these states can be inferred from their
decay modes. Since the $D_{sJ}^{*+}(2317)$ decays into two
pseudoscalars it is likely to be a $0^+$ state, though higher spin
cannot be ruled out. Similar reasoning would assign the
$D_{sJ}^{+}(2460)$ as a $1^+$ state. These assignments are
strengthened by the non-observation of the radiative $\gamma D_s$
transition for the $D_{sJ}^{*+}(2317)$ and its observation for the
$D_{sJ}^{+}(2460)$. Belle has confirmed the assignment for the
$D_{sJ}^{+}(2460)$  by measuring the angular distribution in the
$B$ decay channel, shown in Fig.~\ref{Belle_Dss_hel}.

\begin{figure}
\centerline{\epsfig{figure=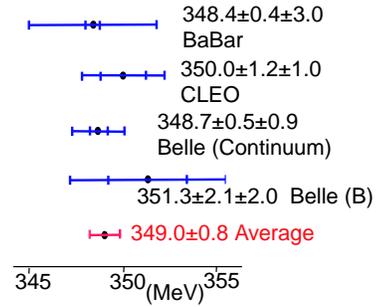,height=1.6in}}
\caption{Measurements of the mass difference between the $D_{sJ}^{*+}(2317)$ and the $D_s^+$.}
\label{2317}       
\end{figure}

Fig.~\ref{2317} summarizes the measurements of the mass difference
between the $D_{sJ}^{*+}(2317)$ and the $D_s^+$. The measurements
are in good agreement. The world average mass difference is
349$\pm$0.8 MeV. Adding the PDG value for the $D_s^+$ mass of
1968.5$\pm$0.6 MeV, we arrive at a mass of 2317.5$\pm$1.0 MeV
\cite{PDG}.

\begin{figure}
\centerline{\epsfig{figure=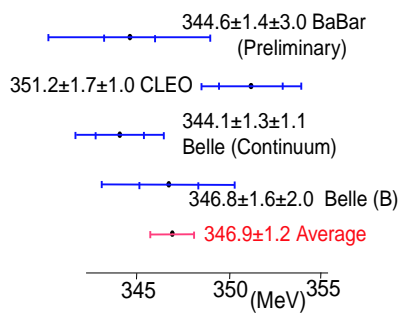,height=1.6in}}
\caption{Measurements of the mass difference between the $D_{sJ}^{+}(2460)$ and the $D_s^{*+}$.}
\label{2460}       
\end{figure}

Fig.~\ref{2460} summarizes the different measurements of the mass
difference for the $D_{sJ}^{+}(2460)$. The CLEO measurement is
somewhat larger than the Belle and BaBar continuum values but
within error. Ultimately mass values using $B$ reconstruction may
be the best way to obtain mass values (for both states) as the
feed across corrections are absent. The world average mass
difference is 346.9$\pm$1.2 MeV. Adding the PDG value for the
$D_s^+$ mass of 2112.4$\pm$0.7 MeV, we arrive at a mass of
2459.3$\pm$1.4 MeV \cite{PDG}. The mass splittings between the
chiral doublets (spin-0 minus spin-1) is $2.1\pm1.4$ MeV,
consistent with zero. This agrees with the ``parity doubling"
predictions using chiral symmetry coupled with HQET
\cite{bardeeneichtenhill} \cite{NRZ}.

CLEO limits the width $\Gamma <$ 7 MeV for both states,
experimentally, although predictions of widths in the quark model
are at the level of ten keV \cite{bardeeneichtenhill}
\cite{Godfrey} \cite{Colangelo}.

Several calculations of the masses of these two narrow states give
close to the  correct values. These include quenched lattice
\cite{Lattice-good}, although an earlier lattice result did not
agree \cite{Lattice-bad}, the MIT bag model \cite{Sadz} and QCD
sum rules using HQET \cite{Dai}. There continues to be some
disagreement, however \cite{Lucha}.

Most properties of these states can be explained if these
particles are $c\bar{s}$ states. The chiral mass splittings
between the $0^+$ and $0^-$ are equal within experimental error to
that between the $1^+$ and $1^-$, as predicted by parity doubling
coupled with HQET \cite{bardeeneichtenhill} \cite{NRZ}. Radiative
decays are present at the expected rate.  One possible exception
is the small branching ratios reported in $B$ decays by Belle.
This has led Browder \etal ~to propose that these states are a
mixture of $c\bar{s}$ and four-quark states \cite{BPP}. Occam's
Razor would imply that more complicated explanations are not
necessary. However, more experimental information coupled with
theoretical ideas will ultimately settle the issue.

\section{Conclusions}

Heavy quark decays is a huge field and I could only supply a short survey here, with many interesting results unfortunately omitted.
Finding the effects of new physics as well as determining CKM parameters often requires the judicious use of theories and models.
Theories should be used when available. Models can be useful and can give us insight into the basic physics. Models, however, when used quantitatively must be checked by comparing with similar processes in order that we can ascertain the errors due to their particular inherent assumptions. These considerations have led me to extract conservative values for $\left|V_{cb}\right|=(424pm 1.2_{exp} \pm 2.3_{thy})\times
10^{-3}$ and $|V_{ub}|=(3.90\pm_{exp} 0.16\pm_{thy} 0.53) \times 10^{-3}.$
These values are then used to fit for CKM parameters in the Standard Model and the allowed region can be compared with measurements of CP violation in $B^{\rm o}\to J/\psi K_s$, for example; that was done by Yamamoto at this conference \cite{Hitoshi}.

Rare decays, first seen in the exclusive channel $B\to K^*\gamma$
and the inclusive channel $b\to s\gamma$ by CLEO have now been
seen in $b\to s\ell^+\ell^-$ and in $B\to K\ell^+\ell^-$ by Belle,
and in $B\to K^*\ell^+\ell^-$ by BaBar and Belle. These channels
can show the effects of new physics when sufficient statistics are
accumulated. The polarization in the $K^*$ mode is especially
important to study. Rare two-body hadronic decays are becoming
precisely measured in many channels. Analysis of these decays is
becoming more and more interesting.

There have been many surprises in the field of heavy quark
physics. The $b$ lifetime was predicted to be very short, below
$10^{-14}$ s. $B^{\rm o}-\overline{B}^{\rm o}$ mixing was supposed
to be too small to observe. The excited $D_{sJ}$ states, were
``known to be" wide.

If anything is predictable in this field it is that we expect
surprises. Thus, finding the effects of New Physics will not be a
great surprise, we expect to do it! What is not known is the kind
of New Physics we will see.

\begin{acknowledgement}
This work was supported by the U. S. National Science Foundation.
Many people helped by providing me with data and discussions. I am
grateful to them all, although it should be clear that inclusion
in this list does not mean they necessarily agree with my
arguments. I thank: M. Artuso, W. Bardeen, T. Barnes, M. Beneke,
M. Bona, J. Butler, K. Ecklund, E. Eichten, L. Gibbons, C. Hill,
A. Kronfeld, U. Langenegger, Z. Ligeti, H. Lipkin, M. Mangano, M.
Neubert, P. Roudeau, J. Rosner,  N. Uraltsev, J. Urheim, M. Wise
and J. C. Wang.
\end{acknowledgement}


%

\end{document}